\documentclass[colordvi,bibtex.titlepage]{elsart} 
\journal{Nuclear Physics B}
\usepackage{amssymb}
\usepackage{amsmath}
\usepackage{graphicx}
\usepackage{dcolumn} 
\usepackage{bm}
\usepackage{longtable}
\usepackage{color}

\definecolor{lightred}{rgb}{1,0.75,0.75}
\definecolor{lightgreen}{rgb}{0.8,1,0.8}

\def\groupargonne{\address[1]
{Physics Division, Argonne National Laboratory, Argonne, Illinois 60439-4843, USA}}
\def\groupbari{\address[2]
{Istituto Nazionale di Fisica Nucleare, Sezione di Bari, 70124 Bari, Italy}}
\def\groupbeijing{\address[3]
{School of Physics, Peking University, Beijing 100871, China}}
\def\groupchina{\address[4]
{Department of Modern Physics, University of Science and Technology of China, Hefei, Anhui 230026, China}}
\def\groupcolorado{\address[5]
{Nuclear Physics Laboratory, University of Colorado, Boulder, Colorado 80309-0390, USA}}
\def\groupdesy{\address[6]{DESY, 22603 Hamburg, Germany}}
\def\groupzeuthen{\address[7]{DESY, 15738 Zeuthen, Germany}}
\def\groupdubna{\address[8]
{Joint Institute for Nuclear Research, 141980 Dubna, Russia}}
\def\grouperlangen{\address[9]{Physikalisches Institut, Universit\"at Erlangen-N\"urnberg, 91058 Erlangen, Germany}}
\def\groupferrara{\address[10]
{Istituto Nazionale di Fisica Nucleare, Sezione di Ferrara and Dipartimento di Fisica, Universit\`a di Ferrara, 44100 Ferrara, Italy}}
\def\groupfrascati{\address[11]
{Istituto Nazionale di Fisica Nucleare, Laboratori Nazionali di Frascati, 00044 Frascati, Italy}}
\def\groupgent{\address[12]
{Department of Subatomic and Radiation Physics, University of Gent, 9000 Gent, Belgium}}
\def\groupgiessen{\address[13]
{Physikalisches Institut, Universit\"at Gie{\ss}en, 35392 Gie{\ss}en, Germany}}
\def\groupglasgow{\address[14]
{Department of Physics and Astronomy, University of Glasgow, Glasgow G12 8QQ, United Kingdom}}
\def\groupillinois{\address[15]
{Department of Physics, University of Illinois, Urbana, Illinois 61801-3080, USA}}
\def\groupmichigan{\address[16]
{Randall Laboratory of Physics, University of Michigan, Ann Arbor, Michigan 48109-1040, USA }}
\def\groupmoscow{\address[17]
{Lebedev Physical Institute, 117924 Moscow, Russia}}
\def\groupnikhef{\address[18]
{Nationaal Instituut voor Kernfysica en Hoge-Energiefysica (NIKHEF), 1009 DB Amsterdam, The Netherlands}}
\def\groupstpetersburg{\address[19]
{Petersburg Nuclear Physics Institute, St. Petersburg, Gatchina, 188350 Russia}}
\def\groupprotvino{\address[20]
{Institute for High Energy Physics, Protvino, Moscow region, 142281 Russia}}
\def\groupregensburg{\address[21]
{Institut f\"ur Theoretische Physik, Universit\"at Regensburg, 93040 Regensburg, Germany}}
\def\grouprome{\address[22]
{Istituto Nazionale di Fisica Nucleare, Sezione Roma 1, Gruppo Sanit\`a and Physics Laboratory, Istituto Superiore di Sanit\`a, 00161 Roma, Italy}}
\def\grouptriumf{\address[23]
{TRIUMF, Vancouver, British Columbia V6T 2A3, Canada}}
\def\grouptokyo{\address[24]
{Department of Physics, Tokyo Institute of Technology, Tokyo 152, Japan}}
\def\groupamsterdam{\address[25]
{Department of Physics and Astronomy, Vrije Universiteit, 1081 HV Amsterdam, The Netherlands}}
\def\groupwarsaw{\address[26]
{Andrzej Soltan Institute for Nuclear Studies, 00-689 Warsaw, Poland}}
\def\groupyerevan{\address[27]
{Yerevan Physics Institute, 375036 Yerevan, Armenia}}

\begin{document}
\begin{frontmatter}
\title {\hfill 
Hadronization in semi-inclusive deep-inelastic scattering on nuclei}
\author[16]{A.~Airapetian},
\author[27]{N.~Akopov},
\author[27]{Z.~Akopov},
\author[7]{E.C.~Aschenauer},
\author[26]{W.~Augustyniak},
\author[27]{R.~Avakian},
\author[27]{A.~Avetissian},
\author[11]{E.~Avetissian},
\author[11]{N.~Bianchi},
\author[18,25]{H.P.~Blok},
% ...
\\[0.25cm]

\centerline{The HERMES Collaboration} 
\groupargonne
\groupbari
\groupbeijing
\groupchina
\groupcolorado
\groupdesy
\groupzeuthen
\groupdubna
\grouperlangen
\groupferrara
\groupfrascati
\groupgent
\groupgiessen
\groupglasgow
\groupillinois
\groupmichigan
\groupmoscow
\groupnikhef
\groupstpetersburg
\groupprotvino
\groupregensburg
\grouprome
\grouptriumf
\grouptokyo
\groupamsterdam
\groupwarsaw
\groupyerevan

\date{\today}
\thanks[now1]{Present address: 36 Mizzen Circle, Hampton, Virginia 23664, USA}

\vskip 1cm

\begin{abstract}
A series of semi-inclusive deep-inelastic scattering measurements
on deuterium, helium, neon, krypton, and xenon targets has been 
performed in order to study hadronization.
The data were collected with the HERMES detector
at the DESY laboratory using a 27.6 GeV positron or electron beam.
Hadron multiplicities on nucleus $A$ relative to those on the deuteron,
$R_A^h$, are presented for various hadrons
($\pi^+$, $\pi^-$, $\pi^0$, $K^+$,  $K^-$, $p$, and $\bar{p}$)
as a function of the virtual-photon energy $\nu$,
the fraction $z$ of this energy transferred to the hadron,
the photon virtuality $Q^2$, and the hadron transverse
momentum squared $p_t^2$.
The data reveal a systematic decrease of $R_A^h$
with the mass number $A$ for each hadron type $h$.
Furthermore, $R_A^h$ increases (decreases) with increasing values of
$\nu$ ($z$), increases slightly with increasing $Q^2$, and is almost
independent of $p_t^2$, except at large values of $p_t^2$.
For pions two-dimensional distributions also are presented.
These indicate that the dependences of $R_A^{\pi}$ on $\nu$ and $z$
can largely be described as a dependence on a single variable $L_c$,
which is a combination of $\nu$ and $z$.
The dependence on $L_c$ suggests in which kinematic conditions partonic
and hadronic mechanisms may be dominant.
The behaviour of $R_A^{\pi}$ at large $p_t^2$ constitutes
tentative evidence for a partonic energy-loss mechanism.
The $A$-dependence of $R_A^h$ is investigated as a function of
$\nu$, $z$, and of $L_c$. It approximately follows an $A^{\alpha}$ form
with $\alpha \approx 0.5 - 0.6$.

\end{abstract}
\end{frontmatter}

\maketitle

\section{Introduction}
\label{sec:intro}

After decades of extensive study, understanding the confinement
of quarks and gluons in hadrons still is one of the great
challenges in hadronic physics. To uncover its nature, hadronic
reactions in a nuclear medium, either cold or hot, are studied.
Typical examples are the measurements of hadron production on
nuclear targets in semi-inclusive deep-inelastic scattering of
leptons~\cite{osborn,ashm,adams,herm1,herm2} and the jet-quenching and
parton energy-loss phenomena observed in ultra-relativistic
heavy-ion collisions~\cite{RHIC1,RHIC2}.
In each case hadron yields are observed that are different from
those observed in the corresponding reactions on free nucleons.

The process that leads from the partons produced in the
elementary interaction to the hadrons observed experimentally
is commonly referred to as hadronization or fragmentation.
According to theoretical estimates the hadronization process occurs
over length scales varying from less than a femtometer to several
tens of femtometers. At these length scales the magnitude of the
strong coupling constant is such that perturbative techniques
cannot be applied. Hence, hadronization is an intrinsically
non-perturbative QCD process, for which only approximate theoretical
approaches are presently available. Experimental data are vital for
supporting these theoretical developments, since they can be used to
gauge or guide the calculations.

The hadronization process in free space has been studied extensively
in $e^+\,e^-$ annihilation experiments \cite{eplemin}. As a result the
spectra of particles produced and their kinematic dependences are
rather well known. However, little is known about the space-time
evolution of the process. Semi-inclusive production of hadrons in
deep-inelastic scattering of leptons from atomic nuclei provides a
way to investigate this space-time development. Leptoproduction of
hadrons has the virtue that the energy and the momentum of the
struck parton are well determined, as they are tagged by the
scattered lepton. By using nuclei of increasing size one can
investigate the time development of hadronization. If hadronization
occurs quickly, i.e., if the hadrons are produced at small distances
compared to the size of atomic nuclei, the
relevant interactions in the nuclear environment involve well-known
hadronic cross sections such as the ones for pion-nucleon
interactions. If, in contrast, hadronization occurs over large
distances, the relevant interactions are partonic and involve the
emission of gluons and quark-antiquark pairs. The two mechanisms
lead to different predictions for the decrease in hadron yield,
known as attenuation, on nuclei as compared to that on free
nucleons.

Most likely, a combination of these two mechanisms contributes
to the observed attenuation of hadron yields on nuclei. This
expectation has led to a range of phenomenological approaches, which
are briefly summarized in the next section. The available
calculations cover a range of possible mechanisms (partonic
energy loss or hadronic absorption) and time (length) scales (from
less than 1 fm to more than 10 fm), which results in different
dependences on the various kinematic variables. In order to
distinguish between these calculations precise hadron attenuation
data are needed as a function of several kinematic variables for a
range of nuclei and for several hadron types.

Exploratory measurements were first performed at SLAC~\cite{osborn}
and later by the EMC~\cite{ashm} and E665~\cite{adams}
collaborations.
More recently many more data have been collected by the HERMES
collaboration at DESY and the CLAS collaboration at the Thomas
Jefferson National Accelerator Facility~\cite{jlab1}.
Some of the HERMES data have been published already~\cite{herm1,herm2}.
The CLAS data are presently being analysed~\cite{jlab2}.
In this paper we present the full results from HERMES on the
multiplicities for the production of pions, kaons, protons, and
antiprotons on helium, neon, krypton, and xenon targets relative to
those for deuterium.
It goes beyond the scope of the present paper to compare the data to
the available theoretical calculations. Instead, prominent features of the
data are identified and used to address two key issues in the study of hadronization:
(i) what are the time or length scales of the process, and
(ii) what are the mechanisms that lead to nuclear attenuation?

The paper is organized as follows. In section~\ref{sec:theory} the
theoretical framework is described, and some representative
theoretical models are summarized. In section~\ref{sec:exper} those
aspects of the HERMES experiment that are relevant to the present
measurements are presented. In section~\ref{sec:anal} the data
analysis is discussed, including the corrections to the raw data and
the systematic uncertainties. The results for the attenuation as a
function of various kinematic variables are presented and discussed
in section~\ref{sec:res}. This section has several subsections in
which the features of the data, especially those related to the
relevant time scales and mechanisms, and the $A$-dependence, are
discussed separately. The results are summarized in the last
section, which also lists the conclusions.


\section{Theoretical framework}
\label{sec:theory}

In order to put the experimental results into perspective,
in this section the concepts that are used in the study of hadronization
are presented, and models that have been developed to describe the
experimental results are briefly discussed.
It is emphasized that the latter is meant only to illustrate
potential interpretations of the data. The experimental data
and the features that they exhibit are the genuine subject.
 
\begin{figure}[!t]
\center{
\includegraphics[width=6cm]{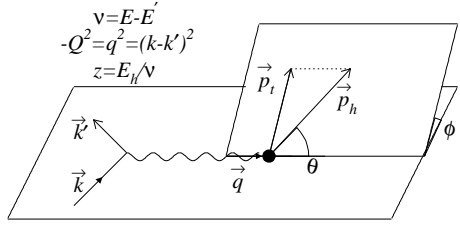}
\caption{Kinematic planes for hadron production in semi-inclusive deep-inelastic
scattering and definitions of the relevant lepton and hadron variables.
The quantities $k=(E,\vec{k})$ and $k'=(E',\vec{k}')$ are the four-momenta of
the incident and scattered lepton, $(E_h,\vec{p}_h)$ is the four-momentum
of the produced hadron, and $\vec{p}_t$ is the transverse momentum
of the hadron.}
\label{kinematicplot}
}
\end{figure}

The hadronization process in a nuclear medium can be studied
by means of semi-inclusive deep-inelastic scattering (SIDIS)
of electrons or positrons from nuclei.
For that purpose the multiplicity ratio $R^h_A$ is introduced,
which is defined as the ratio of the
number of hadrons $h$ produced per deep-inelastic scattering (DIS)
event on a nuclear target with mass number $A$ to that for a
deuterium (D) target.
Figure~\ref{kinematicplot} illustrates the definition of the
relevant lepton and hadron kinematic variables.
The ratio $R_A^{h}$  depends  on the leptonic variables $\nu$,
the energy of the virtual photon, and $Q^2$,
the negative of the four-momentum of the virtual photon squared,
and on the hadronic variables $z = E_h/\nu$, the fraction of the
virtual-photon energy carried by the hadron,
and $p_t^2$, the square of the hadron momentum component transverse
to the direction of the virtual photon.
Thus $R_A^{h}$ can be written as:
\begin{equation}
R_A^h(\nu,Q^2,z,p_t^2)=
\frac{\big(\frac{N^h(\nu,Q^2,z,p_t^2)}{N^e(\nu,Q^2)}\big)_A}
{\big(\frac{N^h(\nu,Q^2,z,p_t^2)}{N^e(\nu,Q^2)}\big)_D},
\label{eq:rformula}
\end{equation}
with $N^h(\nu,Q^2,z,p_t^2)$ the number of semi-inclusive
hadrons at given $(\nu,Q^2,z,p_t^2)$,
and $N^e(\nu,Q^2)$ the number of inclusive DIS leptons
at $(\nu,Q^2)$. Implicit in this definition is the integration
over the angle $\phi$ between the lepton scattering plane and
the hadron production plane (see Fig.~\ref{kinematicplot}).

Experiments at large values of $\nu$~\cite{ashm,adams} give values
$R_A^{h} \approx 1.0$ within the experimental uncertainty. This is interpreted
as an indication that nuclear effects are negligible in that region.
At lower values of $\nu$ the value of $R_A^{h}$ has been found to be
well below unity~\cite{osborn,herm1,herm2}.

Even though hadronization is not yet quantitatively understood, it is generally
assumed that the following processes play a role in leptoproduction of hadrons.
\newline \noindent
After a quark in a nucleon has absorbed the virtual photon,
it can lose energy by scattering from other quarks and by radiating gluons.
As a result the value of $R_A^h$ may be influenced. A change in $R_A^h$ can also
result from the quark in a nucleus having a different distribution
function as a result of partonic rescaling.
These two effects will be called partonic mechanisms.

After a certain time a colourless object, called a prehadron, is formed, which
has the quantum numbers of the final hadron, but not yet its full wave function.
The concept of colour transparency~\cite{CT} is closely related to this.
The prehadron then evolves over some time into the physical hadron.
In the Lund model~\cite{anders} the process of (pre)hadron formation
is described as the building of a colour field (string) between the struck
quark and the residual system. The string gets stretched and breaks into
smaller pieces, each with an (anti-)quark at both ends.
The Lund model gives predictions (see,~e.g.,~Ref.~\cite{bialas3})
for the time $t_c$, and the corresponding (constituent or formation)
length $l_c$, that it takes for the prehadron to be formed,
and the time $t_h$ required for the final physical hadron to be
formed.\footnote{In the literature the name 'formation time or length'
has sometimes different meanings and also different symbols are used. 
Here the name 'formation length' ($l_c$) is used for what is often called
the prehadron (or constituent) formation length, and 'hadron formation
length'~($l_h$) for the hadron (or yo-yo) formation length
(see, e.g., Refs.~\cite{bialas3,kope01}).} 

The average value of the formation length $l_c$, denoted by $L_c$,
is given in the Lund model by
\begin{equation}
L_c = f(z)\frac{\nu}{\kappa},
\label{eq:lc}
\end{equation}
where $f(z)$ is a function that goes as $1-z$ for large values of $z$
(this behaviour is also found in other models, and is due to the fact
that in the limit $z \rightarrow 1$ the struck quark cannot have
lost any energy when producing the final hadron),
and has a broad maximum of about 0.4 around $z=0.35$. The quantity
$\kappa$ is the string tension, which reflects the energy loss of
the leading quark per unit length, usually taken as $\kappa=1.0$~GeV/fm.
Thus, at values of $\nu$ of $5-25$ GeV, the prehadron formation
takes place over distances of about $1-10$ fm, comparable to
the size of a nucleus.
The values of $l_h$ are larger than $l_c$, so the final hadron is often
formed outside the nucleus.

If the (pre)hadron is formed inside the nucleus, it can experience hadronic
interactions, generally called final-state interactions (FSI).
For discussing the effects of these, we will not discriminate between
prehadrons and hadrons, and just talk about hadrons and hadronic effects,
although the relevant cross sections may be different for hadrons and prehadrons.
The first effect of FSI may be the rescattering of the hadron from the nucleons
in the nucleus, the hadron losing energy and possibly 'generating' other, mainly
low energy, hadrons (if these are mesons, the word generate can be taken
literally; if these are nucleons, generated means that they are now in a
continuum state). The effect is thus a loss of hadrons at a given value of
$z$, and an increase of the same hadron, but possibly also of other hadrons,
at lower $z$.
Another effect of FSI can be that the original hadron is absorbed. This is
usually accompanied by the emission of other, again mainly low-energy, hadrons.
So also in this case there is a loss of hadrons at given $z$, and an increase
of other hadrons at lower $z$.
For a full description of all these effects,  coupled-channel calculations
should be performed. However, since these are rather complicated, 
in models usually  only absorption is considered. Clearly this neglects 
the generation of hadrons at lower values of $z$.

The change from partonic mechanisms to hadronic mechanisms, which depends
on the formation length relative to the size of the nucleus, influences $R_A^h$.
Thus the use of various nuclear targets allows one to investigate
the space-time development of the hadronization process.

At present, reliable QCD calculations of quark hadronization
(fragmentation) can not yet be performed because of
the major role of 'soft', i.e., non-perturbative processes.
For that reason various types of models have been developed.

Phenomenological models~\cite{bialas1,gyul,czyz,akopov1} use 
formation times/lengths and absorption cross sections for
the various hadrons in the nuclear medium. Various formulae
for the formation lengths have been used, and in the more advanced
versions two length scales, $l_c$ and $l_h$, are distinguished, as
well as different absorption cross sections for the quark, the prehadron,
and the final hadron. The absorption cross sections are usually
adjusted to obtain the best description of the $\nu$- and $z$-dependence
of the experimental data.

Other (QCD-inspired) models focus on the energy loss that the struck quark
experiences in the nuclear environment~\cite{zakh1,guo01,wang01,arleo}. 
In Refs.~\cite{guo01,wang01} twist-4 contributions to the fragmentation
functions resulting from multiple scattering and gluon
bremsstrahlung in a nuclear medium are calculated. 
A nuclear attenuation proportional to $A^{2/3}$ is predicted,
where the power $2/3$ results from the coherence of the
gluon radiation process~\cite{landau}, which gives
an induced radiative energy loss of a quark traversing a length $L$
of matter proportional to $L^2$~\cite{doksh}.
No hadron absorption is included, as it is assumed that the
hadron is formed outside the nucleus.

The effect of a finite formation length is included in Ref.~\cite{arleo},
in which fragmentation functions are calculated that are again
modified due to partonic energy loss in the nuclear medium
during the time $t_c$.
By using the quark energy loss determined from Drell-Yan data, a reasonable
agreement with existing data is found.
In order to keep the approach as simple as possible, absorption
of the produced (pre)hadron is not taken into account.

Another class of models includes (pre)hadron absorption, with or
without a description of what happens in the hadronization process.
In Ref.~\cite{accardi} $R_A^h$ is described in terms of medium-modified
fragmentation functions supplemented by nuclear absorption.
A parton-rescaling model that has also been used to describe the EMC effect
is used to describe the nuclear modification of the fragmentation functions.
The (average) formation length is taken from the Lund model~\cite{anders}.
Ref.~\cite{kope03} calculates the nuclear attenuation of the (leading) hadron
with $z>0.5$ by including as major ingredients the formation
length $l_c$ and an absorption cross section of the prehadron.
The effect of quark energy loss is found to be small.
In Ref.~\cite{falter1} the hadron attenuation is investigated
within the framework of the Boltzmann-Uehling-Uhlenbeck 
transport model. Since this is a coupled-channel approach,
hadrons are not only absorbed, but can also be produced.
Some choices for the formation time $t_c$, including taking it
to be zero, are studied. 

The theoretical calculations have been compared to the data from 
Refs.~\cite{ashm,herm1,herm2}.
Notwithstanding their different and sometimes orthogonal ingredients,
all models reproduce the global features of the data.
In the case of the $\nu$-dependence, which is best described, the 
reason may be that the decreasing attenuation with increasing
value of $\nu$ is largely due to a simple increase of the formation time
$t_c$ with $\nu$ in the target rest frame due to Lorentz dilatation.
The dependence of $R_A^h$ on $z$ and $A$, which in general is less well
described, may be more discriminating, especially when more detailed
data are available.

\section{Experiment}
\label{sec:exper}
The measurements were performed with the HERMES spectrometer~\cite{hermesdetector}
using a 27.6~GeV positron or electron beam stored in the HERA ring at DESY.
Some data were collected using a 12.0~GeV positron beam~\cite{nat02,lag05},
but since the amount of data was much less and only pions or all hadrons
were identified, they are not included here.
Typical beam currents were 40~mA down to 5~mA.
The spectrometer consists of two mirror-symmetric halves, located above and
below the lepton beam pipe. A flux-exclusion plate in the
midplane of the magnet prevents deflection of the lepton (and proton)
beams passing through the center of the spectrometer.
The scattered leptons and the produced hadrons were detected within an angular
acceptance of $\pm$ 170 mrad horizontally and $\pm$ (40 -- 140) mrad vertically. 
The lepton trigger was formed by a coincidence between signals from three
scintillator hodoscope planes and a lead-glass calorimeter.
A minimum energy deposit in the latter of 3.5 GeV (1.4 GeV) 
for unpolarized (polarized) target runs was required.

The data were collected during the years 1999, 2000, 2004, and 2005, using
unpolarized nuclear (He, Ne, Kr, Xe) and polarized or unpolarized deuterium (D)
gaseous targets internal to the storage ring (see Table~\ref{tab:Table1}).
In 1997 also data on nitrogen were taken, but since at that time
the RICH detector (see below) was not yet available, only data for
pions and all hadrons together could be presented~\cite{herm1}.
Therefore these data are not included here.
The yields from polarized deuterium were summed over the two spin orientations. 
The target gases were injected into a 40~cm long tubular open-ended storage cell.
Using an Unpolarized Gas Feed System~\cite{shin} it is possible to provide
D, He, N, Ne, Kr, or Xe targets with relatively  high areal densities
(between 10$^{14}$ and 10$^{17}$ nucleons/cm$^2$),
resulting in luminosities ranging from 10$^{31}$ to 10$^{33}$cm$^{-2}$s$^{-1}$.
Such high-density runs were taken at the end of HERA fills, with typical
currents of 15 to 5 mA and beam lifetimes of one hour.
This made it possible to accumulate the data for these
targets in only a few days of integrated beam time.
The luminosity was measured using elastic scattering of the beam particles
from the electrons in the target gas: Bhabha scattering
for a positron beam, M{\o}ller scattering for an electron
beam~\cite{benish}. 
\begin{table}[htb]
\begin{center}
\caption{Overview of the HERMES nuclear attenuation measurements.}
\label{tab:Table1}
\vskip 0.3cm
\begin{tabular}{|l|c|l|l|l|l|l|r|} \hline\hline
Year&$E$ (GeV)&Target& Identified hadrons&\hspace*{2.5em} Ref.\\
\hline\hline
 1997 & 27.6 & D, N & $h^{\pm}$, $\pi^{\pm}$ & \cite{herm1}, \cite{herm2} \\
 1999 & 27.6 & D, Kr &
$h^{\pm}$, $\pi^{\pm}, \pi^0$, $K^{\pm}$, $p$, $\overline{p}$ & \cite{herm2}, this work
\\
 2000 & 27.6 & D, He, Ne& $\pi^{\pm}, \pi^0$, $K^{\pm}$, $p$, $\overline{p}$ &
this work \\
 2000 & 12.0 & D, N, Kr &$h^{\pm}$, $\pi^{\pm}$ & \cite{nat02} \\
 2004 & 27.6 & D, Kr, Xe
 & $\pi^{\pm}$, $K^{\pm}, \pi^0$, $p$, $\overline{p}$ & this work \\
 2005 & 27.6 & D, Kr, Xe
 & $\pi^{\pm}$, $K^{\pm}, \pi^0$, $p$, $\overline{p}$ & this work \\
\hline\hline
\end{tabular}
\end{center}
\end{table}
There are several particle identification (PID) detectors in the HERMES
spectrometer.  Details on the performance and use of these PID detectors can be
found in Ref.~\cite{longdelq}.
Electrons and positrons are identified by combining the information
from a lead-glass calorimeter, a scintillator hodoscope preceded by
two radiation lengths of lead (the pre-shower detector), and a
transition-radiation detector.

The identification of charged pions, kaons, protons, and antiprotons
is accomplished using the information from the
Ring-Imaging \v{C}erenkov detector (RICH)~\cite{RICH},
which replaces a threshold  \v{C}erenkov counter used in the previously
reported measurements on nitrogen~\cite{herm1}. 
The RICH detector uses two radiators,  a 5 cm thick wall of silica-aerogel
tiles followed by a large volume of C$_4$F$_{10}$ gas, to provide
separation of pions, kaons, and (anti)protons.
Together these provide good particle identification for charged hadrons
in the momentum range $2<p<15$~GeV, with limited contamination from
misidentified hadrons.
The identification efficiencies and contaminations for pions, kaons,
protons, and antiprotons were determined in
a Monte Carlo simulation as a function of the hadron momentum and
multiplicity in the relevant detector half.
These performance parameters were verified in a limited kinematic
domain using known particle species from identified resonance decays.
They were used in a matrix method to unfold the true hadron
distributions from the measured ones.
Systematic uncertainties in the unfolding were estimated by using matrices
determined in different ways, see Refs.~\cite{RICH,RICHwww} for details.

The electromagnetic calorimeter~\cite{calo} provides neutral pion
identification through the detection of two clusters without an
accompanying track, originating  from the two decay photons.

\section{Data Analysis}
\label{sec:anal}

The analysis procedure is similar to the one described in detail in
Refs.~\cite{herm1,herm2}, where the nitrogen and first krypton data
were presented. Since the publication of Ref.~\cite{herm2}, more data
on krypton were taken and all data were analysed in a wider kinematic range.

The hadron multiplicity ratio $R_A^h$ as defined in Eq.~\ref{eq:rformula} was
determined as a function of the leptonic ($Q^2$ and $\nu$) and hadronic ($z$
and $p_t^2$) variables for all identified particles and all targets.
The kinematic constraints imposed on the scattered leptons were identical
for all analysed data:
$Q^2>1$~GeV$^2$, $W=\sqrt{2M\nu+M^2-Q^2} >2$~GeV (where $M$ is the nucleon mass)
for the invariant mass of the photon-nucleon system,
and $y=\nu/E<0.85$ for the energy fraction of the virtual photon.
The constraints on $W$ and $y$ were applied in order to exclude nucleon
resonances and to limit the magnitude of the radiative corrections to
$R_A^h$, respectively. The resulting value of $x_{Bj}=Q^2/2M\nu$ ranged from
0.023 to 0.8.

As mentioned in the previous section, charged hadrons were identified in
the momentum range $2.0 - 15.0$~GeV by using the RICH detector.
For the identification of the neutral pions through the detection of their
decay photons, each of the two photon clusters
was required to have an energy of at least 0.8~GeV.
The background was evaluated in each kinematic bin by fitting the two-photon
invariant mass spectrum with a Gaussian plus a polynomial that
fits  the shape of the background due to uncorrelated photons.
The number of  detected neutral pions was obtained by integrating the peak,
corrected for background,
over a $\pm$2$\sigma$ range with respect to the centroid of the Gaussian.
The low momentum limit for the neutral pions was set at 2.5 GeV
in order to reduce backgrounds.

The integrated luminosities for all years and targets are listed in Table~\ref{tab:Table2}.
Typical numbers of observed DIS leptons and identified hadrons are listed
in Table~\ref{tab:Table3}.

\begin{table}[htb]
\caption{Integrated luminosities (in pb$^{-1}$) for the various data sets.}
\label{tab:Table2}
\begin{center}
\vskip 0.3cm
\begin{tabular}{|c|c|c|c|c|c|} \hline\hline
Target&1999&2000&2004&2005&Sum\\ \hline
D&32.3&119.7&35.7&61.7&249.4 \\
He&&27.9&&&27.9\\
Ne&&84.2&&&84.2\\
Kr&26.1&&29.5&21.1&76.7\\
Xe&&&21.2&21.4&42.6\\
\hline\hline
\end{tabular}
\end{center}   
\end{table}

\begin{table}[htb]
\caption{Number of DIS leptons and identified hadrons 
collected on deuterium and krypton targets in 1999, 2004, and 2005 combined.
The numbers are for the following kinematic constraints:
$Q^2>1$~GeV$^2$, $\nu>6$~GeV, $W>2$~GeV, and
$z>0.2$.  
}
\label{tab:Table3}
\begin{center}
\vskip 0.3cm
\begin{tabular}{|c|c|c|c|c|c|c|c|c|} \hline\hline
Target &\hspace*{0.5em}DIS&$\pi^{+}$&$\pi^{-}$&$\pi^0$&$K^{+}$&$K^{-}$&$p$&$\overline{p}$\\
\hline\hline
D&6669k&706k&575k&232k&146k&62k&131k&23k\\
\hline\hline
Kr&3516k&286k&232k&90k&68k&26k&69k&8k\\
\hline\hline
\end{tabular}
\end{center}
\end{table}

Most of the systematic uncertainties related to the detector, the
reconstruction efficiencies and particle identification practically cancel
in the ratio of the multiplicities.
In determining the multiplicity ratios, deuterium data collected in the same year
as the data for the heavier target were used to avoid uncertainties due to possible
different conditions or functioning of the HERMES spectrometer during the years.
It was verified that the multiplicity ratios obtained in different
years were consistent within the statistical and systematic uncertainties.

The multiplicity ratios were also inspected as a function of the hadron
angles $\theta_x$ and $\theta_y$ with respect to the beam direction
in order to investigate whether the values of $R_A^h$ depend on the
geometrical acceptance of the spectrometer. After applying (small)
corrections for changes in average kinematics with $\theta_x$ or
$\theta_y$, no effect was found beyond the statistical and systematic
uncertainties.
The dependence of the value of $R_A^h$ on the azimuthal angle $\phi$ of the
hadron (see Fig.~\ref{kinematicplot}) was also investigated, since it is known
that the SIDIS cross section on the proton and deuteron depends on this variable.
It was found that $R_A^h$ was constant as a function of $\phi$ within the
statistical and systematic uncertainties.

The data for the multiplicity ratios were corrected for radiative processes
in the manner described in Ref.~\cite{RCRC}.  The code of Ref.~\cite{akush}
was modified to include the measured SIDIS cross sections.
The radiative corrections (RC) were applied to both the inclusive
and the semi-inclusive parts in Eq.~\ref{eq:rformula}.
For the inclusive cross sections elastic, quasi-elastic, and inelastic
processes need to be taken into account, whereas for the semi-inclusive ones
only inelastic radiative processes contribute.
The correction for the ratio of the latter was taken to be independent of $z$.
Since the inelastic radiative effects are almost the same for the nuclei
$A$ and $D$, the size of the radiative corrections applied to $R_A^h$ was 
small over most of the kinematic range.
Only in kinematic regions of DIS where the elastic and quasi-elastic tails
are non-negligible, i.e., at the highest value of $\nu$ and lowest
value of $Q^2$ (low $x_{Bj}$), is there a noticeable effect on $R_A^h$,
with a maximum (increase) of $R_A^h$ of about 7\% for xenon
and krypton, 4.5\% for neon, and 1\% for helium.

Since the usual interpretation of the definition of $R_A^h$
(see Eq.~\ref{eq:rformula}) is that it should only include hadrons formed
in the fragmentation process, a correction has to be made for measured hadrons
that are the decay products of directly produced mesons.
The main effect is on the charged-pion multiplicities as a result of the
decay of exclusively produced $\rho^0$ mesons
(for pions from other mesons and for other hadrons the contribution is small).
That may affect the multiplicities for positive (negative) pions by an amount
ranging from about 1\% at low $z$ up to 30\% (45\%) at high values of $z$ ,
as estimated from a Monte Carlo simulation.
The effect on  the multiplicity ratio $R_A^{\pi}$ is much smaller,
but does not cancel completely since the  $\rho^0$ mesons also
interact with the nuclear medium.
Taking into account the measured nuclear transparency~\cite{CT02}
for $\rho^0$ mesons, the maximum remaining effect on $R_A^{\pi}$,
which occurs for $z$-values of 0.7-0.8, was estimated 
to be about 2(4)\%, 3(5)\%, 3.5(6)\%, and 4(7)\%
in the case of helium, neon, krypton, and xenon, respectively.
The first(second) number applies for positive (negative) pions.
These values were included in the systematic uncertainties.

The total systematic uncertainties include the uncertainties of 
hadron identification (1.5\% for neutral pions,  0.5\%  for
charged pions, 2\% for kaons, 2\% for protons, and  6\%  for antiprotons),
overall efficiency ($<2$\%), $\rho^0$-meson production for positive
(0.3\% - 4\%) and negative (0.3\% - 7\%) pions, and the effects of using
different parameterizations of fragmentation functions and
distribution functions in the RC calculations ($<2$\%).


\section{Results and Discussion}
\label{sec:res}

In this section the experimental results are presented and 
the dependences of the multiplicity ratios $R_A^h$ on the various kinematic
variables and the mass number $A$ of the nucleus are discussed.
Unless specified otherwise the data are shown with the following constraints:
$\nu>6.0$ GeV, $z>0.2$, and $x_F>0$, where $x_F$ is given by
\begin{equation}
x_F= p^{*}_\parallel/ p^{*max}_\parallel ,
\end{equation}
with $ p^{*}_\parallel$ being the component of the hadron momentum
in the direction of the momentum transfer in the
virtual-photon nucleon center-of-mass system.
Together with that on $z$, the constraint on $x_F$ will
 reduce possible contributions from target fragmentation.

\subsection {Multiplicity ratio for identified hadrons} 
\begin{figure*}[!ht]
\center{
\includegraphics[width=14cm]{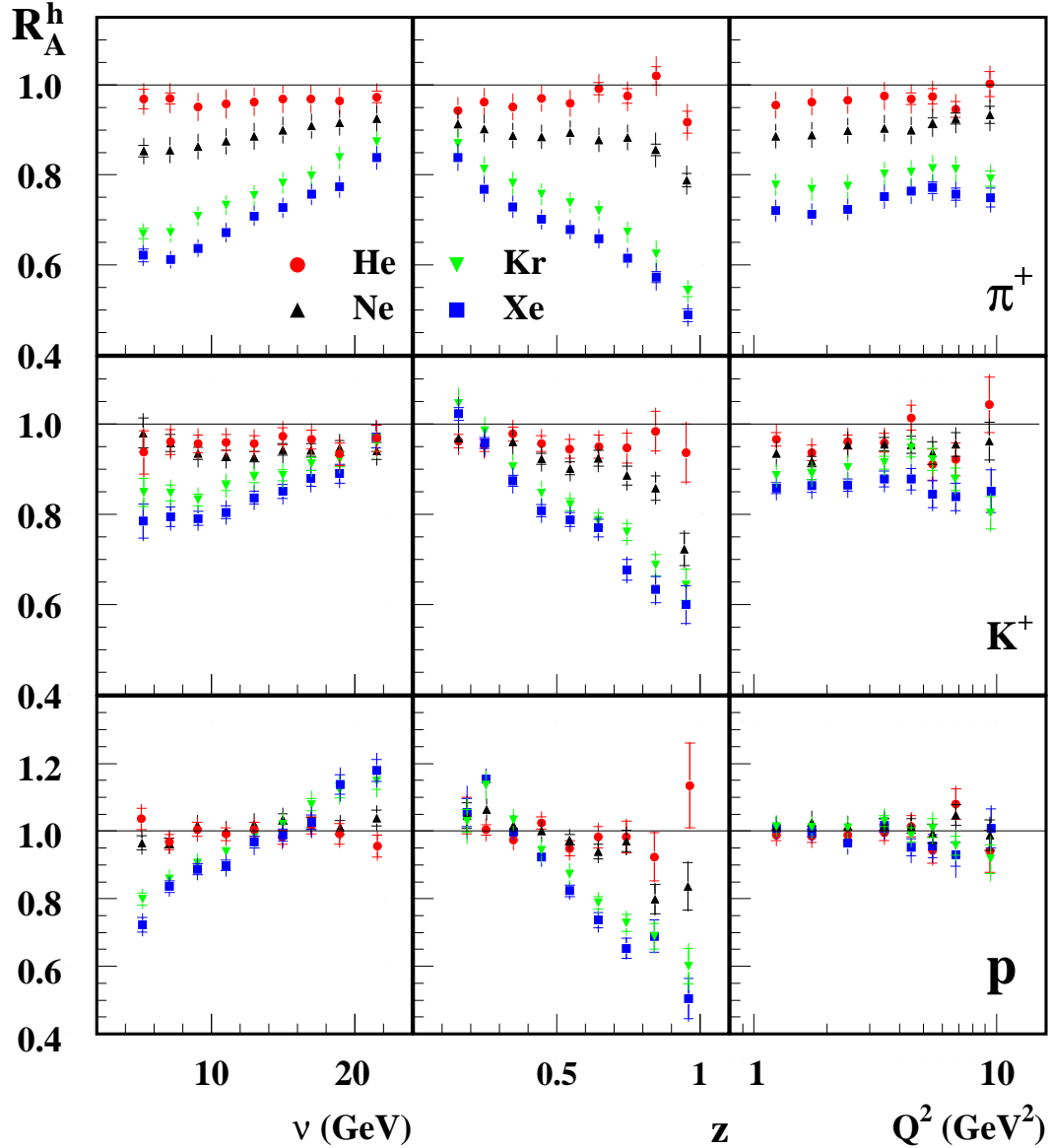}  
\caption{Values of $R_A^h$ for positively charged hadrons as a function
of $\nu$, $z$, and $Q^2$. 
The data as a function of $\nu$ are shown for $\nu>4$ GeV and
those as a function of $z$ for $z>0.1$.
The inner error bars represent the statistical uncertainty, while the outer
ones show the total uncertainty.}
\label{fig:rplus}
}
\end{figure*}

\begin{figure*}[!ht]
\center{
\includegraphics[width=14cm]{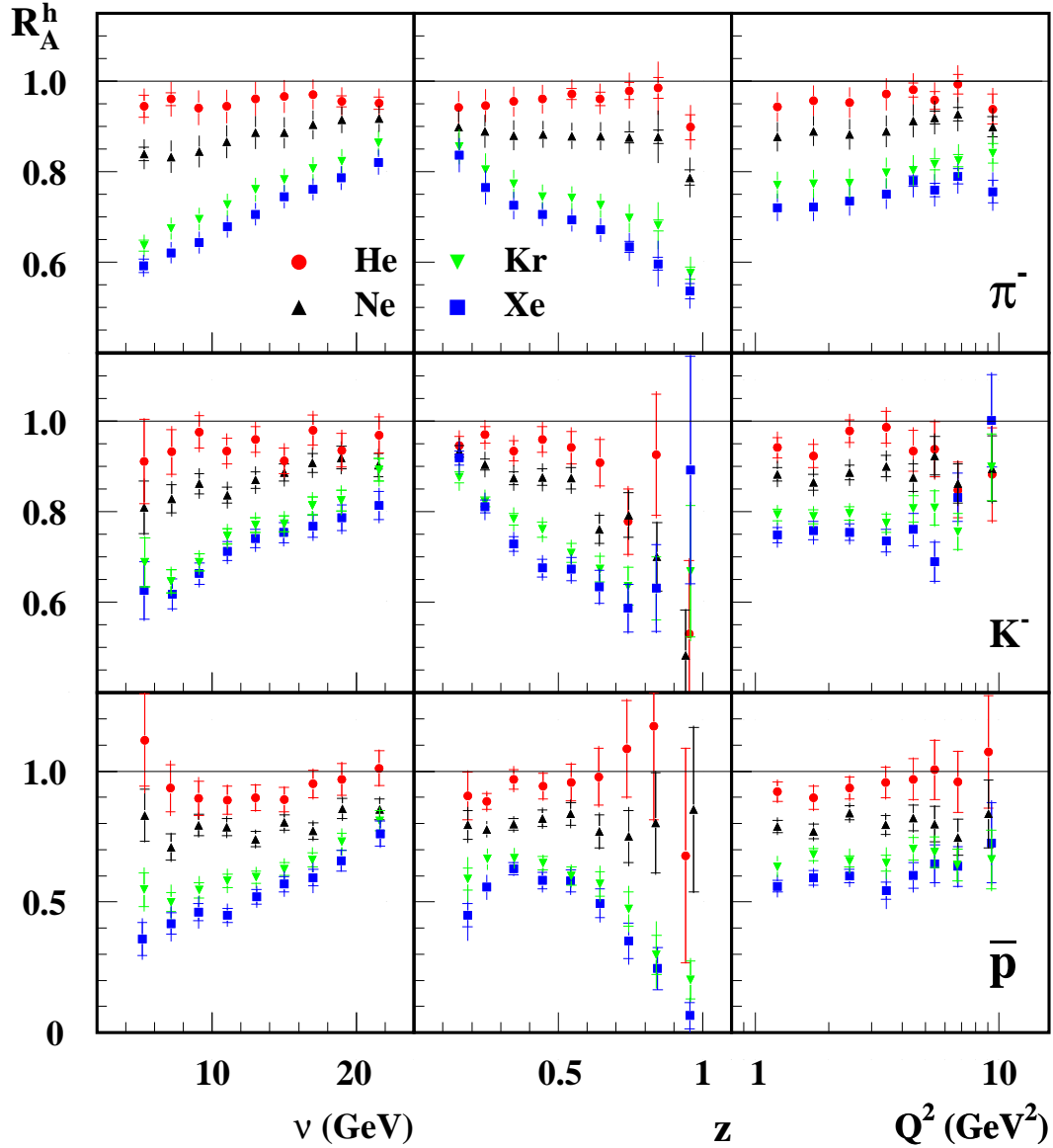}
\caption{Values of $R_A^h$ for negatively charged hadrons as a function
of $\nu$, $z$, and $Q^2$. 
The data as a function of $\nu$ are shown for $\nu>4$ GeV and
those as a function of $z$ for $z>0.1$.
Error bars as in Fig.~\ref{fig:rplus}.  }
\label{fig:rmin}
}
\end{figure*}

\begin{figure*}[!ht]
\center{
\includegraphics[width=14cm]{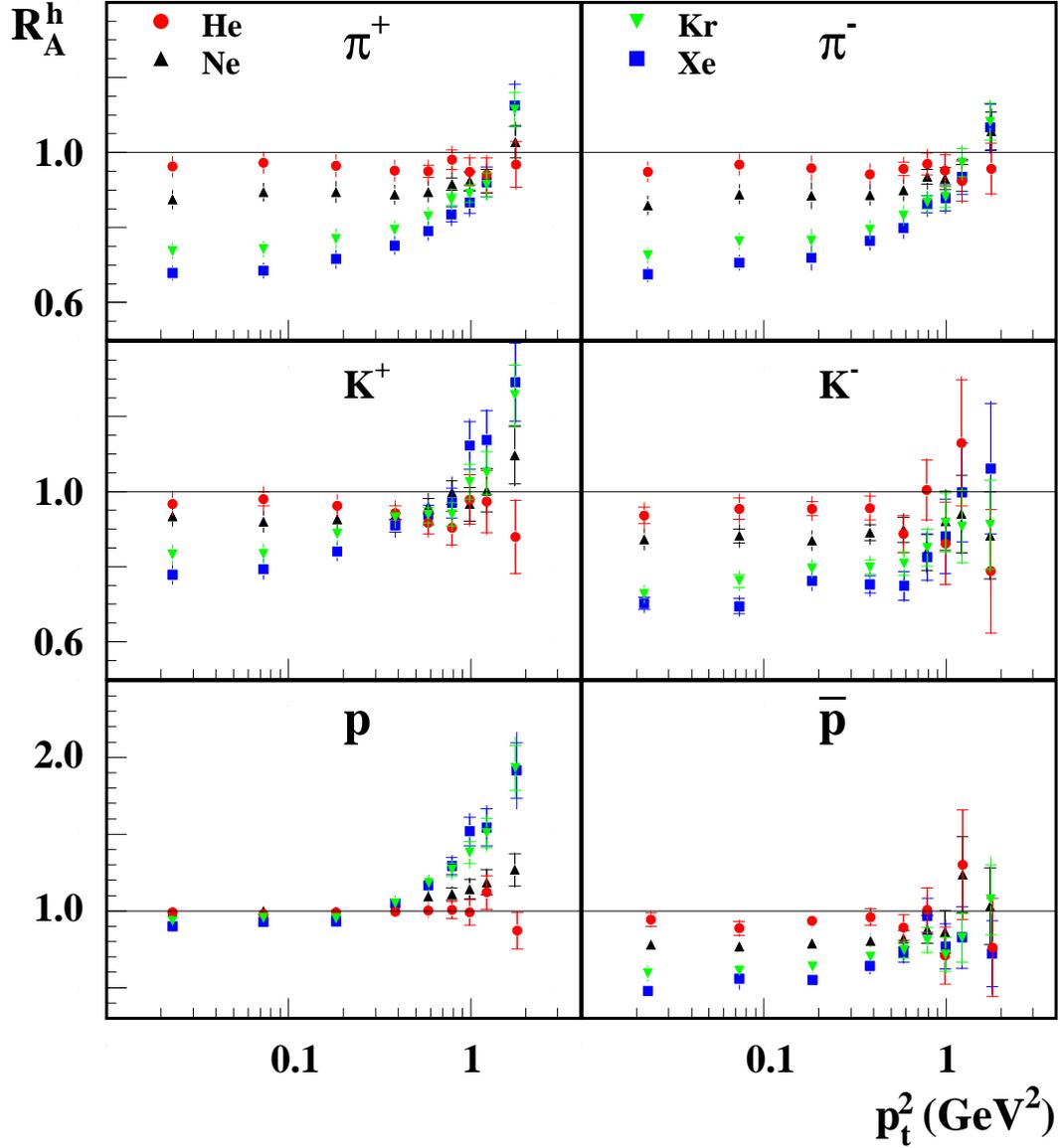}
\caption{Values of $R_A^h$ for positively (left panel) and negatively
(right panel) charged hadrons as a function of $p_t^2$.
Error bars as in Fig.~\ref{fig:rplus}.  }
\label{fig:ptpn}
}
\end{figure*}

\begin{figure*}[!ht]
\center{
\includegraphics[width=14cm]{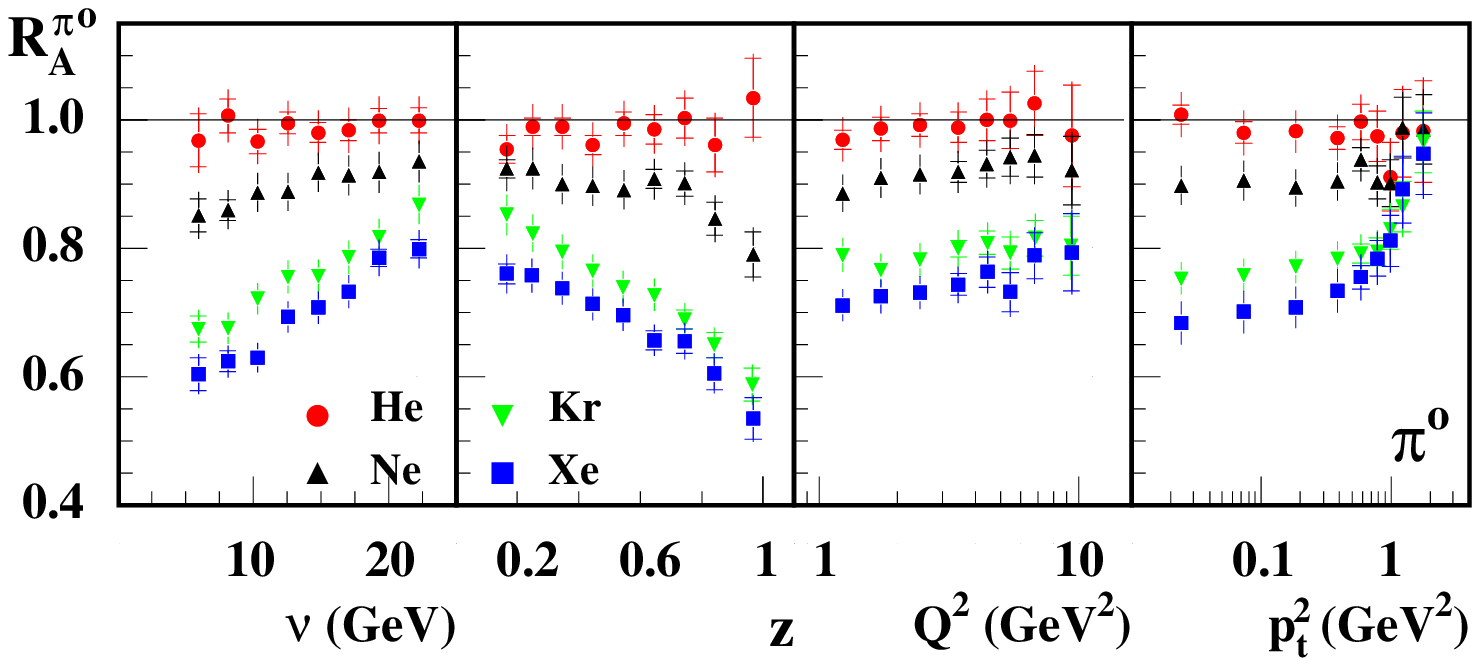} 
\caption{Values of $R_A^h$ for neutral pions as a function of $\nu$,
$z$, $Q^2$, and $p_t^2$.
The data as a function of $z$ are shown for $z>0.1$.
Error bars as in Fig.~\ref{fig:rplus}.  }
\label{fig:rzero}
}
\end{figure*}

Figures~\ref{fig:rplus}-\ref{fig:rzero}
show the dependence of $R_A^h$ on $\nu$, $z$, $Q^2$, and $p_t^2$
for the various nuclei for all identified hadrons:
positively charged (pions, kaons, and protons), negatively charged
(pions, kaons, and anti-protons), and neutral ones (pions).
The inner error bars in these figures represent the statistical uncertainties,
while the outer ones are for the total uncertainty (statistical plus 
systematic, added quadratically). The systematic uncertainty is mainly a
scale uncertainty, affecting the values of $R_A^h$ for the various
values of $\nu$, $z$, $Q^2$, or $p_t^2$ in the same way.

In presenting the results for $R_A^h$ as functions of one of the four
independent variables ($\nu$, $z$, $Q^2$, $p_t^2$) only, $R_A^h$ was
integrated (within the acceptance of the experiment) over the others.
Because of acceptance effects and because in general the dependence of
$R_A^h$ on $\nu$, $z$, $Q^2$, and $p_t^2$ does not factorize 
($R_A^h(\nu,z,Q^2,p_t^2) \neq R_1(\nu)R_2(z)R_3(Q^2)R_4(p_t^2)$),
this integration may introduce false dependences.
This is mainly relevant in case of $\nu$ and $z$, where the average value
of $\nu$ ($z$) changes non-negligibly depending on the value of $z$ ($\nu$).
Table~\ref{tab:avkin} in the Appendix gives an indication of the size of
these correlations by listing the average values of the kinematic quantities
that were integrated over for the various dependences in the case of
produced pions.
All data are available in detail from Ref.~\cite{tabulated_data_database}.

Before discussing in the  following subsections the dependence of $R_A^h$
on the kinematic variables $\nu$, $z$, $Q^2$, and $p_t^2$ in 
detail, first some global features of the data are discussed.

The basic feature of the data is the decrease
of $R_A^h$ with increasing  value of the mass number $A$ of the nucleus.
Qualitatively this is understood as being due to increased partonic
(quark energy loss) or hadronic (absorption) effects.
Furthermore, there is a large similarity between the data for $\pi^+$,
$\pi^-$, and $\pi^0$, and a clear difference  between those for $K^+$ and $K^-$,
and those for $p$ and $\overline{p}$. Also here there are some simple arguments
to explain these features at least qualitatively.

Since we use (almost) isoscalar targets and the production of $\pi^+$ and
$\pi^-$ on protons or neutrons is only slightly different,
both the production and absorption
of pions is in the first instance independent of their charge.
The values of $R_A^h$ for $K^+$ and $K^-$ show a similar behaviour
as a function of the various variables, but $R_A^{K^-}$
is almost everywhere smaller than $R_A^{K^+}$.
A positive kaon can be produced directly from the struck quark (in the
language of string breaking models it is a rank 1 hadron),
but  a negative kaon can only be produced in more complicated
string breakings (rank 2 or higher), except at small values of $x_{Bj}$,
where sea quarks start to play a larger role.
This is reflected in the production rate on deuterium being much larger
for $K^+$ than for $K^-$, see Table~\ref{tab:Table3}, and leads to a steeperr
dependence of the $K^-$ fragmentation function on $z$, and hence a reduced
production if the parton has lost energy before hadronization.
Also, due to their quark content, nuclear absorption cross sections
are larger for $K^-$ than for $K^+$.  Thus both parton energy loss and
absorption of the produced kaon can qualitatively explain
the observed difference between $R_A^{K^+}$ and $R_A^{K^-}$.
However, when comparing the multiplicity ratios for pions and kaons,
it is seen that $R_A^{K^-} \approx R_A^{\pi^-}$, whereas $R_A^{K^+} > R_A^{\pi^+}$.
Given that both pions and $K^+$ particles are rank-1, and $K^-$ rank 2 or higher,
and that nuclear absorption cross sections for both $K^+$ and $K^-$
are smaller than for pions, these features are not readily explained
by the behaviour of fragmentation functions or absorption cross sections.

The results for protons cannot really be related to those for
any of the other particles. Because protons are already present in a nucleus,
an appreciable fraction of them may not come from hadronization.
This is reflected in the very large difference in production of
$p$ and $\overline{p}$ on deuterium, see Table~\ref{tab:Table3}.
Furthermore, as discussed in section~\ref{sec:theory}, in final-state interactions
they generally are not absorbed, but give rise to more nucleons
(both protons and neutrons), thus possibly even increasing $R_A^h$
at lower $z$. 

Antiprotons feature a rather strong attenuation, which might be attributed
to the relatively large $\overline{p}N$ cross section.

\subsubsection {$\nu$-dependence}

The first systematic experimental study of the $\nu$-dependence
of attenuation was reported in Ref.~\cite{ashm}, where the range
$20<\nu<200$~GeV was investigated. It was shown that the nuclear
attenuation decreased with increasing value of $\nu$, and essentially
vanished at $\nu>50$~GeV.
The HERMES data are more informative, since they are in the region of $\nu$
where the attenuation becomes appreciable, include particle identification,
and the statistical precision is much better.

The leftmost columns in
Figs.~\ref{fig:rplus},~\ref{fig:rmin}, and~\ref{fig:rzero} show 
that in almost all cases the attenuation decreases (the value of
$R_A^h$ increases) with increasing values of $\nu$.
(For He this behaviour presumably is present as well, but small 
compared to the uncertainties in the data points.)
In the absorption-type models this is explained as being due to an
increase of the formation length in the rest frame of the nucleus
at higher $\nu$ due to Lorentz dilatation, resulting in a larger
fraction of the hadronization taking place outside the nucleus.
In partonic models the quark energy loss leads effectively to a shift $\Delta z$
in the argument of the fragmentation function, and thus an attenuation that
is proportional to $\epsilon/\nu$, with $\epsilon$ the quark energy loss.

For protons $R_A^h$ increases at higher values of $\nu$ to well above unity
for Kr and Xe. Here the following should be realized.
The value of $\langle z \rangle$ is correlated with $\nu$,
e.g., the value of $\langle z \rangle$ for the lowest $\nu$-bin is
about 0.57, whereas for the highest $\nu$-bin it is 0.35.
Since the value of $R_A^h$ strongly increases with decreasing value of $z$
(see the next subsection), a large fraction of the strong increase at
high $\nu$ is in fact due to the dependence of $R_A^h$ on $z$.
Such an effect may play a role for other particles, e.g., for
$K^+$, too.

\subsubsection {$z$-dependence}
As can be seen from the second column in
Figs.~\ref{fig:rplus},~\ref{fig:rmin}, and~\ref{fig:rzero} for all hadron types
$R_A^h$ is largely constant with $z$ for He, while it decreases with
increasing $z$ for Ne and especially for Xe and Kr.
In parton energy-loss models this results from the strong decrease of
the fragmentation function at large $z$ in combination with the
$\Delta z$ resulting from the energy loss. In absorption-type models
the overall decrease of $R_A^h$ with increasing $z$ is assumed to be due
to a decrease in the formation length in combination with (pre)hadronic absorption.
The increase of $R_A^h$ at large $z$ calculated in Ref.~\cite{arleo} is
not observed in the data.

For the heavier targets $R_A^h$ rises strongly at low $z$.
Presumably this is due to large FSI effects in these nuclei, through which
particles of higher energy lose energy or get absorbed, generating (other)
lower-energy particles.

As in the case of large values of $\nu$, the value of $R_A^h$ for protons
rises above unity at small $z$. This presumably is a result of
large rescattering of protons and other produced particles with protons
in the target. Part of the increase is due to the fact that
the average value of $\nu$ decreases considerably with $z$,
from about 18 GeV for the lowest bin to about 10 GeV for the
highest bin. This explains also why even at the smallest $z$ the value
of $R_A^h$ is still lower than that for the highest value of $\nu$.

Apart from featuring rather small values of $R_A^h$, down to almost zero,
the $z$-dependence of $R_A^h$ for antiprotons is special in that it stays
constant or even decreases slightly at small values of $z$, where
$R_A^h$ for other particles increases strongly for the heavier targets.
This probably can be attributed to the fact that in final-state interactions
the chance that an anti-proton survives, or is produced, is relatively small.
This would support the idea that the rise of $R_A^h$ for other particles
at $z<0.3$ is due to FSI effects and suggests that
the difference in behaviour of anti-protons and the other particles
at small values of $z$ may be a sensitive check on coupled-channel
calculations of FSI effects.

\subsubsection{$Q^2$-dependence}
The rightmost column of Figs.~\ref{fig:rplus},~\ref{fig:rmin}, and ~\ref{fig:rzero}
shows for pions a small $Q^2$-dependence, which is slightly stronger for
the heavier nuclei.
For kaons and (anti-)protons no $Q^2$-dependence is discernable. 
Hence, the attenuation is not very sensitive to $Q^2$, which
means that integrating over $Q^2$ when studying other
dependences does not introduce false dependences. 

In the twist-4 energy-loss model of Refs.~\cite{guo01,wang01}
a $Q^2$-dependence of $R_A^h$ of the form ${R \sim a \ln Q^2}$
is found, which is consistent with the data. 
In Ref.~\cite{kope03} the calculated  $Q^2$-dependence is the result of two 
counteracting processes, which results in a rather small $Q^2$-dependence
that is larger for Kr than for Ne, in global agreement with the data.
When describing the attenuation purely as the result of a modification
of the effective fragmentation function~\cite{arleo},
a slight increase of $R_A^h$ with $Q^2$ is predicted.
The deconfinement model~\cite{accardi} predicts for all nuclei a slight decrease
of $R_A^h$ with $Q^2$, which is not supported by the data. 

\subsubsection{$p_t^2$-dependence}
Figure~\ref{fig:ptpn} and the rightmost column of Fig.~\ref{fig:rzero}
show for the heavier nuclei a rise of $R_A^h$ at high $p_t^2$.
Such an effect was first observed by EMC~\cite{ashm} for all charged hadrons
taken together, but has been measured now for separate identified hadrons.
The phenomenon is also known from heavy-ion collisions, where it is
referred to as the Cronin effect~\cite{CR75}.
Compared to experiments with heavy ions, the use of a lepton probe
has the advantage that initial-state interactions do not play a role,
except for shadowing effects, which are small in the $x_{Bj}$ range of the
present experiment.
The observed rise at high $p_t^2$ is attributed to a broadening of the
$p_t^2$ distribution.  In principle this can result from partonic rescattering
as well as from hadronic final-state interactions.
In the next subsection this will be looked at in  more detail in the case of pions.

\subsection {Two-dimensional multiplicity ratio for pions}

\begin{figure*}[!ht]
\center{
\includegraphics[width=14cm]{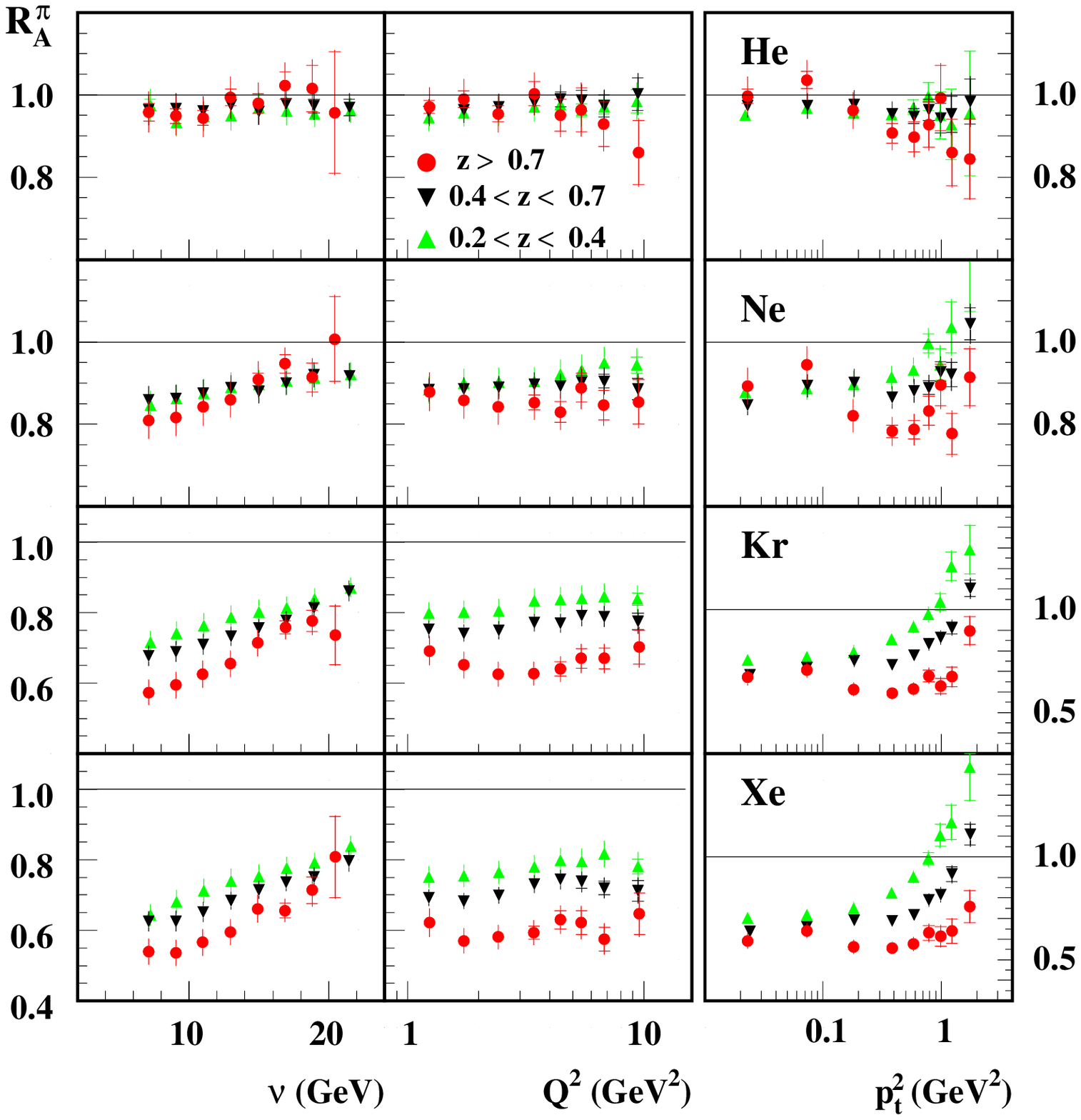}  
\caption{Values of $R_A^h$ for charged pions for three $z$ ranges.
Error bars as in Fig.~\ref{fig:rplus}.}
\label{fig:3difz}
}
\end{figure*}

\begin{figure*}[!ht]
\center{
\includegraphics[width=14cm]{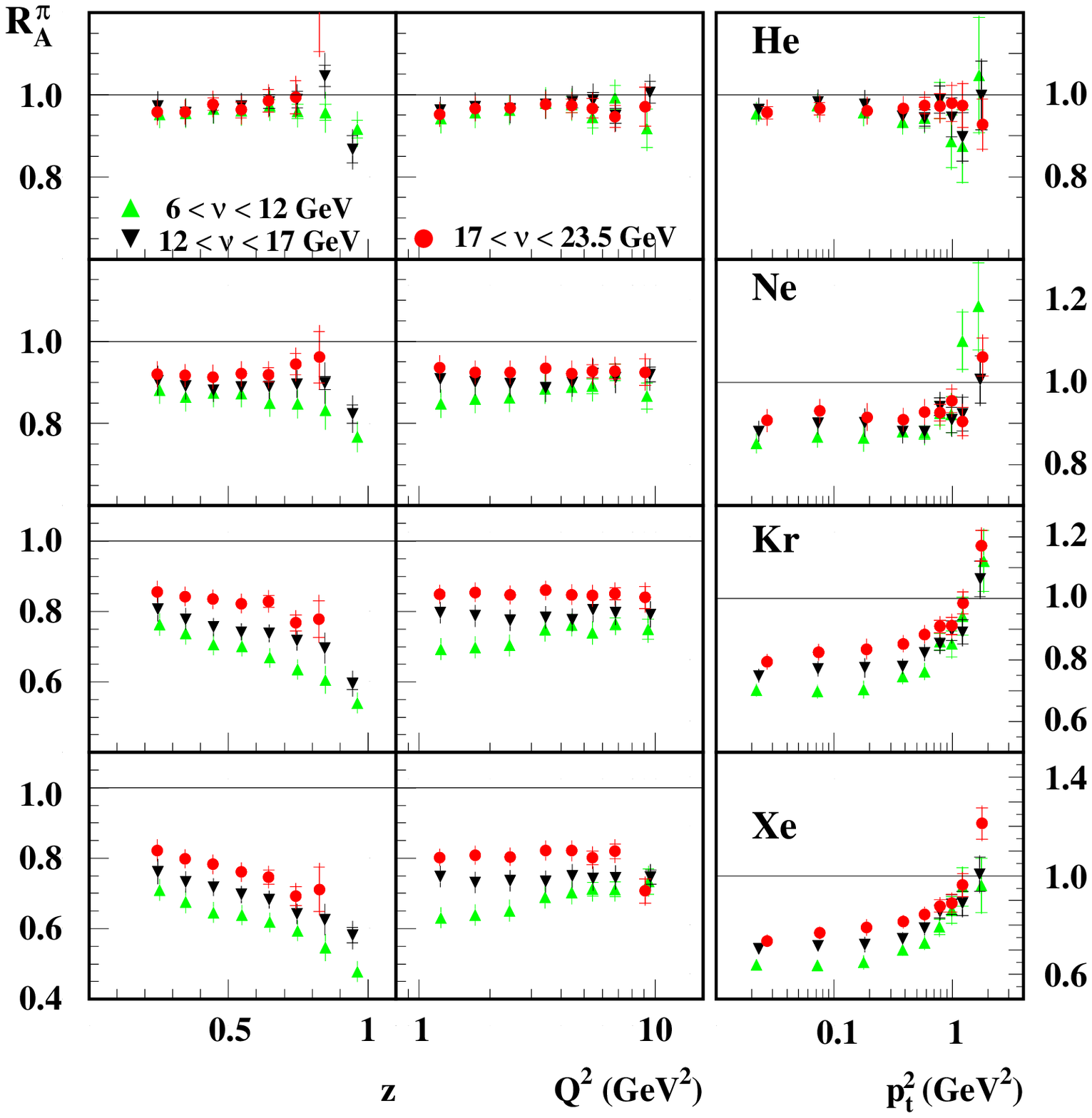}  
\caption{Values of $R_A^h$ for charged pions for three $\nu$ ranges.
Error bars as in Fig.~\ref{fig:rplus}.}
\label{fig:3difnu}
}
\end{figure*}

\begin{figure*}[!ht]
\center{
\includegraphics[width=14cm]{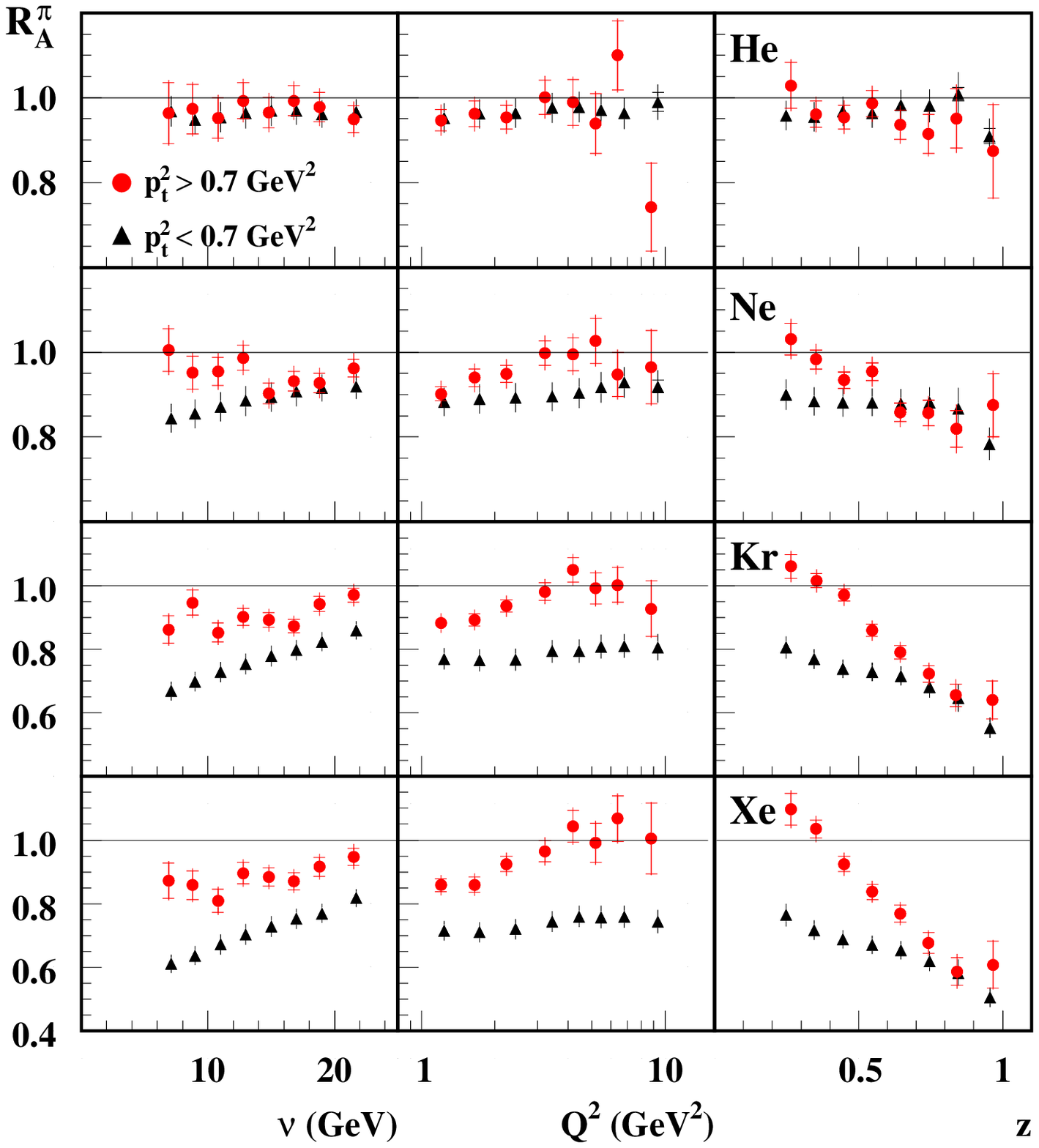}  
\caption{Values of $R_A^h$ for charged pions for two $p_t^2$ ranges.
Error bars as in Fig.~\ref{fig:rplus}.}
\label{fig:2difpt}
}
\end{figure*}

In order to investigate the behaviour of $R_A^h$ in more detail, the data for
$\pi^+$ and $\pi^-$ production, which have the best statistical precision
and are consistent within uncertainties, were combined for
a two-dimensional binning, see Figs.~\ref{fig:3difz}-\ref{fig:2difpt}.
The bins used are $0.2-0.4-0.7-1.2$ for $z$, $6.0-12.0-17.0-23.5$ GeV for $\nu$, 
and smaller or larger than 0.7~GeV$^2$ for $p_t^2$.
This has the added advantage that the correlation between, e.g., the 
average values of $\nu$ and $z$ mentioned in relation to the one-dimensional
distributions, is strongly reduced.

Figure~\ref{fig:3difz} shows the dependence of $R_A^{\pi}$ on
$\nu$, $Q^2$, and $p_t^2$ for three bins in $z$. The left hand column
indicates that the dependence on $\nu$ hardly depends on $z$.
The $Q^2$-dependence is similar for the various $z$-bins
(Monte-Carlo simulations show that the rise of $R_A^{\pi}$ at the
lowest $Q^2$ for the highest $z$ range is due to a relatively
large contribution of pions coming from $\rho^0$ decay).
Therefore, the dependence on $z$ is not affected when integrating over $Q^2$.
The data in the rightmost column indicate that the increase of $R_A^h$
for Kr and Xe at large $p_t^2$ is smaller for larger $z$. Such a
$z$-dependence of the $p_t^2$-dependence was predicted in Ref.~\cite{kope03}.
The points for the highest $z$ range show a bump at small values of $p_t^2$.
This is due to a relatively large contribution
of pions coming from $\rho^0$ decay in this part of the phase space.


Figure~\ref{fig:3difnu} shows the dependence of $R_A^{\pi}$ on
$z$, $Q^2$, and $p_t^2$ for three bins in $\nu$.
The second and third column indicate that the dependence of $R_A^h$ on $Q^2$ and
$p_t^2$ depends only weakly on $\nu$, which allows one to integrate over
these variables without introducing spurious correlations.
The leftmost column shows for He and Ne (and perhaps for Kr)
some interesting features in that
there seems to be a change of the $z$-dependence with the value of $\nu$,
$R_A^{\pi}$ first being about constant or even rising slightly with $z$,
and then dropping, the turnover point occurring
at lower $z$ for the lower $\nu$-bin.
This behaviour is studied in more detail in the next subsection.

Figure~\ref{fig:2difpt} shows the dependence of $R_A^{\pi}$ on
$\nu$, $Q^2$, and $z$ for two bins in $p_t^2$. 
The leftmost column shows that at large $p_t^2$ the values of $R_A^h$ are
larger and that the dependence on $\nu$ largely disappears.
This is clearly correlated with the fact that $R_A^h$ increases at
large $p_t^2$ (Cronin effect), as discussed in the previous subsection,
and indicates that this effect is largely independent of $\nu$.
The rightmost column of this figure shows that the Cronin effect
disappears at high $z$.
This is at least consistent with the idea that the rise of $R_A^h$ at large $p_t^2$
(broadening of the $p_t^2$-distribution) is of partonic origin. In the limit
$z \rightarrow 1$ there is no room for partonic rescattering, because the
parton is not allowed to have any energy loss (see, e.g.,~Ref.~\cite{kope03}).
In principle, rescattering of the produced (pre)hadron could lead to
the observed behaviour, too, but since the rescattering cross sections are
relatively small, the data discussed suggest a partonic mechanism.
At the same time, this tells that the attenuation in the limit $z \rightarrow 1$
is due purely to a hadronic absorption mechanism.

\subsection {Dependence of $R_A^{\pi}$ on formation length}
\begin{figure}[!ht]
\center{
\includegraphics[width=12cm]{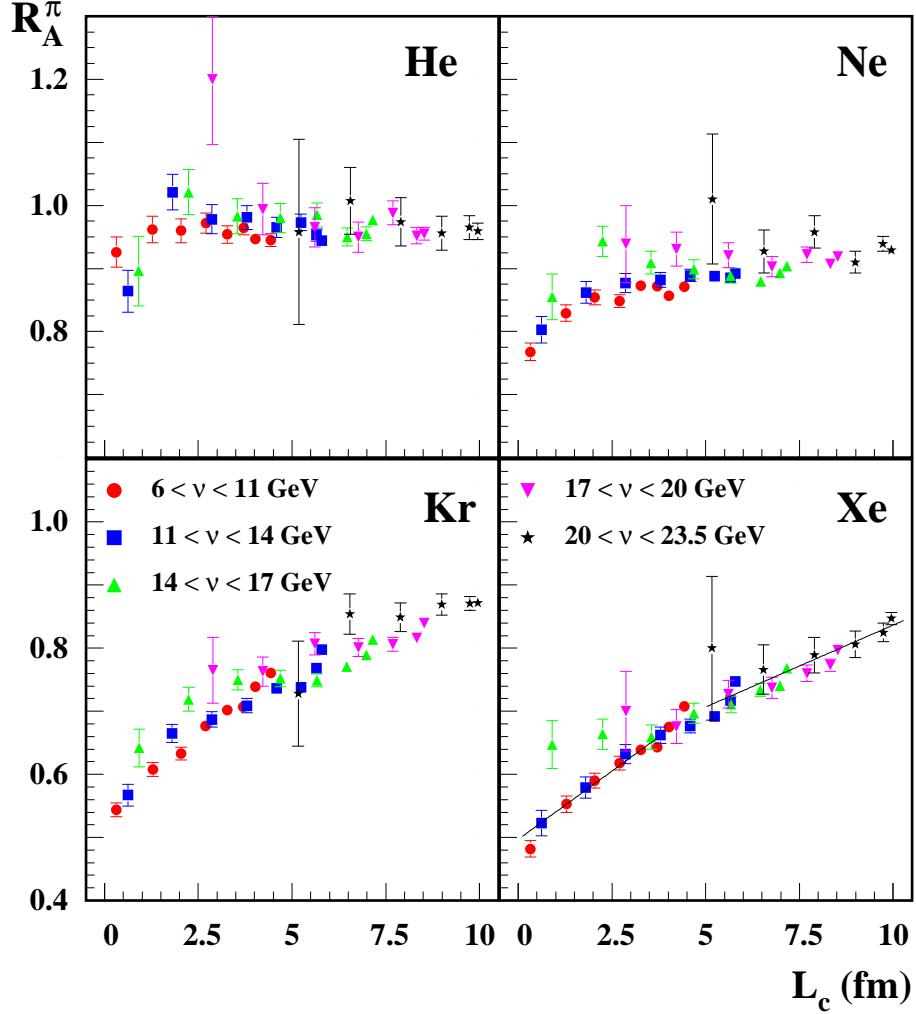}
\caption{Values of $R_A^{\pi}(\nu,z)$ for He, Kr, Ne, and Xe as a
function of the variable $L_c$, see Eqs.~\ref{eq:lc} and \ref{eq:fz}.
The various $\nu$-bins are indicated by different symbols.
Within the same $\nu$ bin the $z$ bins used are
$0.2-0.3-0.4-0.5-0.6-0.7-0.8-0.9-1.0$, and the value of $z$
decreases from left to right. In order not to complicate the figure,
only statistical error bars are shown. The systematic uncertainty
is mainly a scale uncertainty of about 3\%.}
\label{fig:Rnuzlc}
}
\end{figure}

Given the ideas of how hadronization proceeds in time, as~e.g.,~in the Lund
model, the formation length, which depends on both $\nu$ and $z$,
may be a more efficient variable for describing the kinematic
dependence of $R_A^h$ than $\nu$ and $z$. 
This idea was recently pursued in Ref.~\cite{acc06}.
In order to investigate this, values of $R_A^{\pi}$ versus $L_c$
for various values of $\nu$ and $z$ are shown in Fig.~\ref{fig:Rnuzlc}.
Here Eq.~\ref{eq:lc} is used with
\begin{equation}
\label{eq:fz}
f(z) = z^{0.35}(1-z)
\end{equation}
and $\kappa=1$~GeV/fm. 
This form for $f(z)$ is a convenient parametrization obeying the constraints
at $z \rightarrow 0$ and $z \rightarrow 1$, and gives values for $L_c$ as a
function of $z$ closely resembling the ones obtained with the Lund model.

A clear correlation can be observed between the values
of $R_A^h$ and $L_c$, with only a relatively small residual
spread at any fixed value of $L_c$.
Evidently most of the dependence of $R_A^h$ on $\nu$ or $z$ in
Figs.~\ref{fig:3difz}-\ref{fig:3difnu} can be described as
a dependence on $L_c$, which thus acts as a scaling variable.
Upon close inspection it is seen that for almost all $\nu$ values
the data points for the lowest two $z$ values (the two rightmost
in the sequence of same symbols) bend slightly upwards.
This behaviour is most likely due to large rescattering effects
in the lowest $z$-bin (note the relatively strong rise of $R_A^{\pi}$
at low $z$ in Figs.~\ref{fig:rplus} and \ref{fig:rmin}).

For He the value of $R_A^{\pi}$ rises for the first few points at
small $L_c$ and then becomes constant.
For Ne the initial rise extends over more points, and changes
then to a much more gradual one.
For Kr and Xe the change in slope is much more gradual, but
still noticeable. This is illustrated by the two straight lines
in the plot for Xe, which represent fits to the data for the
ranges $L_c<4$~fm and $L_c>5$~fm. 
This suggests the following interpretation:
at the larger values of $L_c$, which are (much) larger than the
size of these nuclei\footnote
{It should be realized that the average distance that
a created parton travels through a nucleus (assuming it is not
absorbed) is only $\frac{3}{4} R$, with $R$ the radius of that nucleus,
because the virtual photon can interact anywhere in a nucleus.
Thus, even for Kr with a radius of about 5 fm,
hadronic mechanisms become small when $L_c>4$ fm.  
}
, even if the absolute scale of $L_c$ may have some uncertainty
because the value of $\kappa$ is not precisely known,
one probably sees a partonic mechanism.
The data for Ne suggest that this mechanism still has a dependence
on $L_c$, presumably through the underlying values of $\nu$ and $z$.
The drop of $R_A^{\pi}$ for low values of $L_c$ then results most probably
from hadronic mechanisms.
The data on Kr and Xe are consistent with this interpretation.
At low values of $L_c$ there is a strong attenuation (stronger
in the larger nucleus Xe) due to hadronic absorption (on top of
the partonic contribution). 
Since the partonic effect must disappear when $z \rightarrow 1$, 
the value of $R_A^{\pi}$ for $L_c \rightarrow 0$ at finite $\nu$
is purely due to the hadronic mechanism. The disappearance of
the partonic contribution in this limit may be an explanation why the
data for the lower values of $L_c$ at given value of $\nu$ tend to lie
above the data at the same $L_c$ with lower $\nu$-value.
At higher values of $L_c$ the influence of absorption becomes smaller, since
the prehadron is increasingly produced outside the nucleus.
Thus above the values of $L_c$ where the slope changes (below 2~fm
for He, and around $L_c=2.5$, 4, and 5~fm for Ne, Kr, and Xe, respectively),
one presumably observes mainly a partonic mechanism.
These tentative conclusions can only be substantiated by model
calculations that include both partonic and hadronic mechanisms,
and that give a good description of the measured $\nu$ and $z$
dependences of $R_A^h$.

\begin{figure}[!ht]
\center{
\includegraphics[width=10cm]{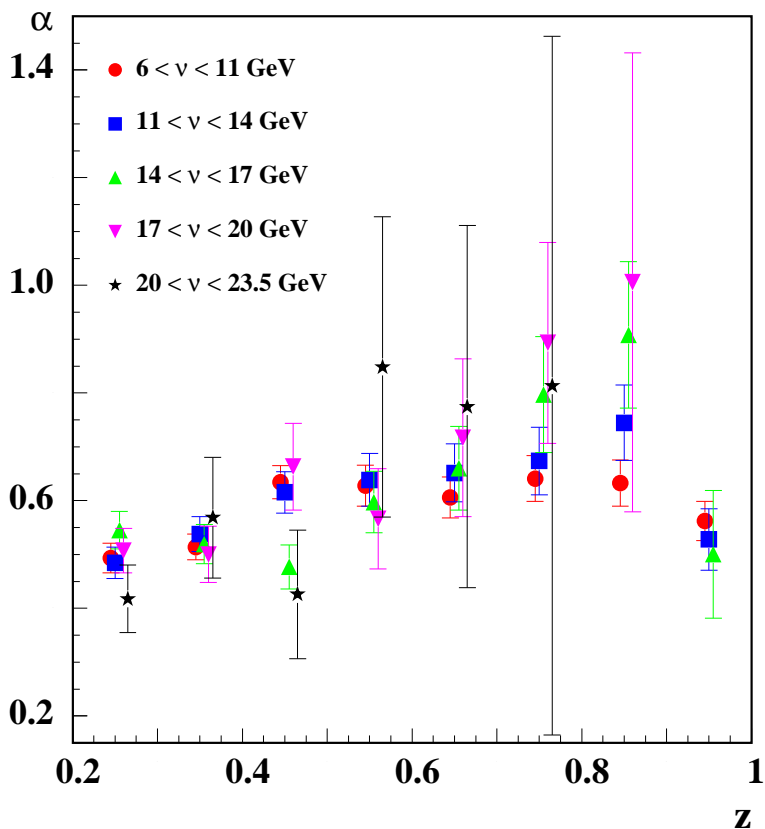}
\caption{\label{fig:albe-nuz} Dependence of the parameter
$\alpha$ (see Eq.~\ref{eq:adep}) on the value of $\nu$ and $z$
for the combined sample of charged pions. Points for different values
of $\nu$ are slightly offset in $z$ for better visibility.}
}
\end{figure}

\begin{figure}[htb]
\center{
\includegraphics[width=10cm]{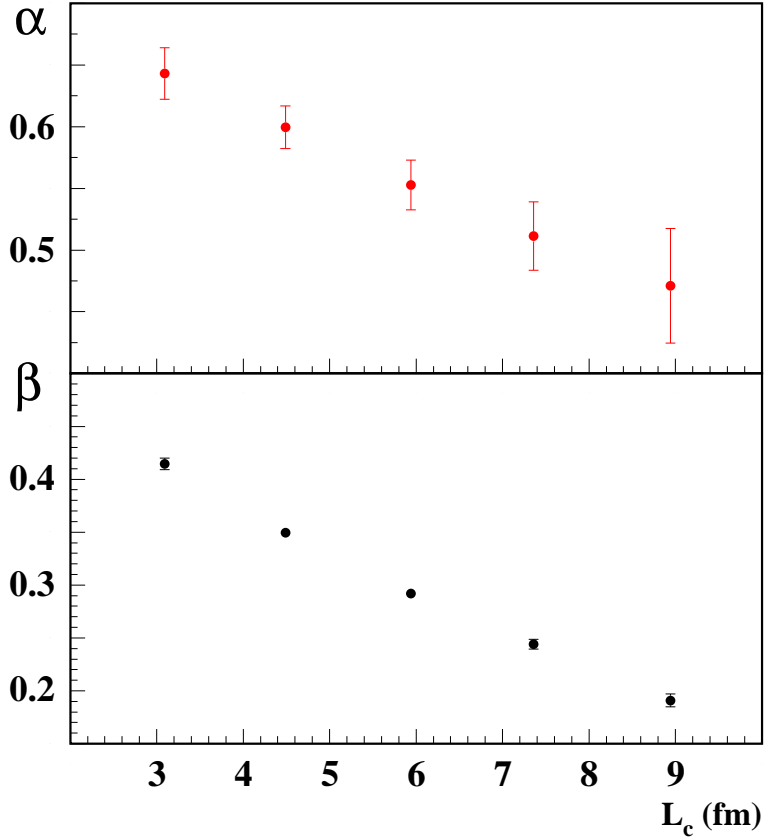}
\caption{\label{fig:albe-lc} The dependence of the parameters
$\alpha$ and $\beta$ on the value of $L_c$ for $0.3<z<0.7$.}
}
\end{figure}

\begin{figure}[htb]
\center{
\includegraphics[width=14cm]{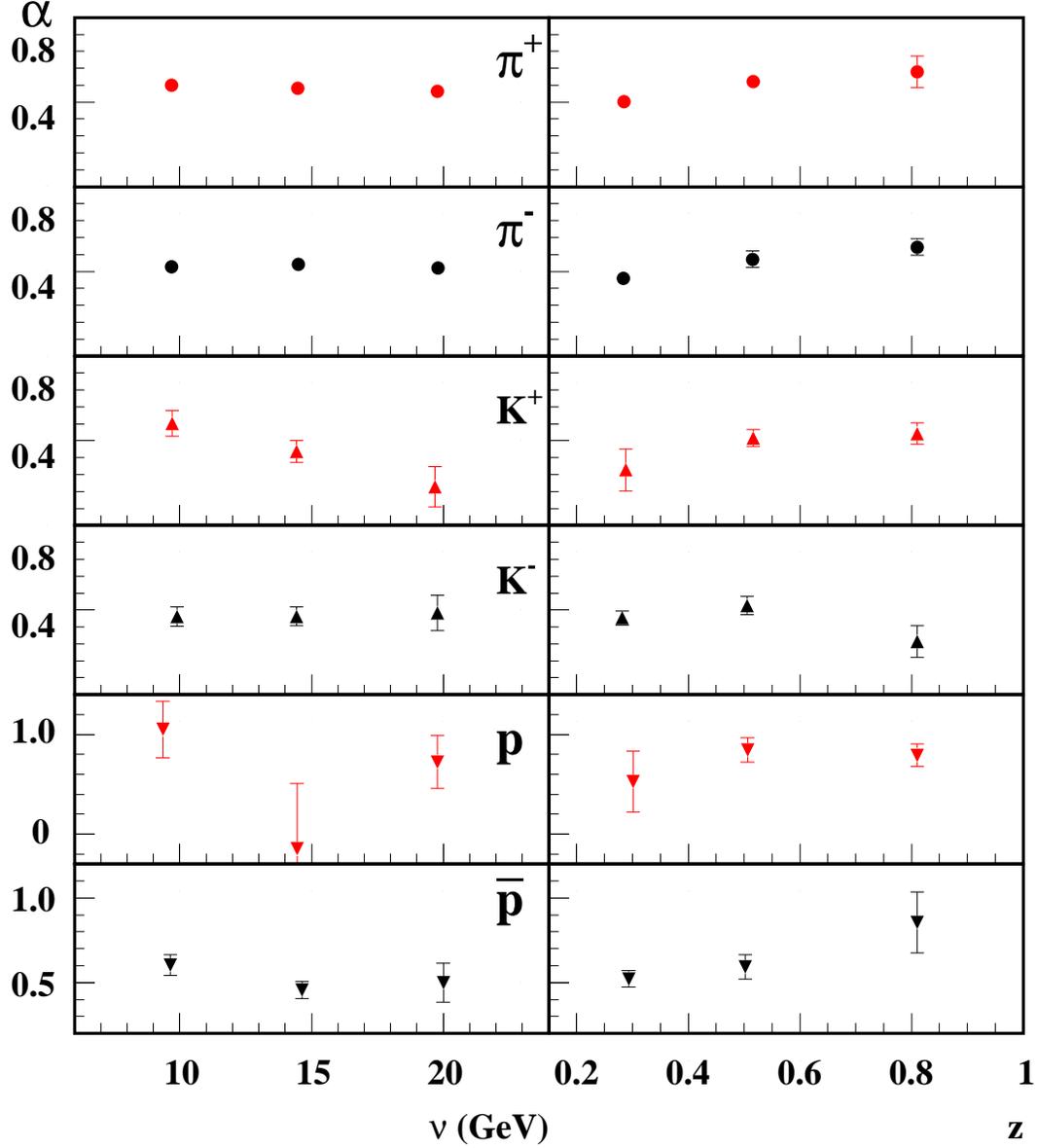}
\caption{\label{fig:albet_gar} Values of the parameter $\alpha$  
for pions, kaons, and (anti-)protons in three $\nu$ ranges
(left six panels), and in three $z$ ranges (right six panels).}}
\end{figure}

\subsection{$A$-dependence}

The $A$-dependence of $R_A^h$ was fitted in a phenomenological way
by using as the fit function
\begin{equation}
\hskip -0.02cm
R_A^h(A) = \exp[-\beta (A/100)^\alpha] .
\label{eq:adep}
\end{equation}
The scale factor of 100 is introduced to reduce the correlation
between $\alpha$ and $\beta$. At the same time the value of $\beta$
is representative now of the attenuation of a nucleus with $A=100$.
For moderate values of $\beta A^\alpha$  this formula is equivalent
to $1-R_A^h = \beta A^\alpha$, a parametrization that
has been used before~\cite{osborn}, but the present form has the advantage
that $R_A^h$ does not become unphysical for very large values of $A$.

However, since $R_A^h$ is the ratio of the multiplicity in nucleus A vs.~the
multiplicity in the deuteron, the fit function of Eq.~\ref{eq:adep} is internally
inconsistent, unless one assumes no attenuation in the deuteron.
Assuming that the attenuation depends on the (average) density times
the radius of the nucleus, in Ref.~\cite{blo06} it is found that
the attenuation in a deuteron can be described with an effective
$A$-value of about~0.6.
The influence of this on the value of $\alpha$ obtained using
Eq.~\ref{eq:adep} is fairly small (about -0.07). Therefore, all results to
be discussed in the following are based on Eq.~\ref{eq:adep},
but in drawing conclusions it should be realized that the real value
of $\alpha$ is probably slightly lower. More information on the performed fits with different type of 
fitting functions can be found in~\cite{zaventhesis}.

A second remark concerns what $A$-dependence is expected from
different types of models for the attenuation.
As already mentioned in section~\ref{sec:theory},
the attenuation in the parton energy-loss model of
Refs.~\cite{guo01,wang01} is given as $1-R_A \sim L^2 \sim A^{2/3}$,
whereas in (Glauber type) absorption models it is often presumed that
$1-R_A \sim L \sim A^{1/3}$. However, these estimates are too simple.
Taking realistic matter distributions of nuclei yields effective values
of $L^2$ that are proportional to $A^{0.74}$~\cite{blo06}.
Furthermore it was demonstrated already in Ref.~\cite{accardi}
that the inclusion of a distribution for the formation length 
in absorption calculations increases the exponent above the value
of 1/3, yielding values for $1-R_A$ that
approximately follow an $A^{2/3}$ pattern for large values of $L_c$,
if the nucleus is described as a sphere with constant density.
Using realistic matter distributions the authors of Ref.~\cite{blo06}
find the exponent of $A$ to be 0.40, 0.54, and 0.60 
when $L_c=0$, 2, and 4~fm, respectively.

In the fits only the statistical uncertainties in the values of $R_A^h$ have
been used, as the systematic uncertainty in the values of $R_A^h$ is largely
a scale uncertainty. The influence of the latter on the value of $\alpha$
was found to be about 0.05.

\subsubsection{Pions}

First the $A$-dependence of pion production was studied as a function
of $\nu$ and $z$, using the data binned in both $z$ and $\nu$.
The results for $\alpha$ as a function of $z$ for the various
$\nu$-bins are shown in Fig.~\ref{fig:albe-nuz}.
The values of $\beta$ (not shown) reflect the global behaviour of $R_A^\pi$
(more attenuation at higher $z$ and lower $\nu$) that is visible
from earlier figures.
There is an increase of $\alpha$ from about 0.5 to 0.6 with $z$ and
possibly at higher $z$ a slight increase with $\nu$. However, in the
latter region the results may be influenced by contributions from
the decay of the $\rho^0$.
The results for the highest $z$-bin do not follow the pattern of
the lower $z$-bins. This is due to the behaviour of $R_A^h$ for especially
He and Ne, seen in Fig.~\ref{fig:3difnu}. As mentioned before,
this could result from the formation length becoming so small that even in
He and Ne hadronic mechanisms become important.

Given these results for the separate $\nu$- and $z$-dependences
it has also been investigated how $\alpha$ and $\beta$ depend
on the value of $L_c$. For that purpose the data were binned
in five $L_c$-bins, using the values of $\nu$ and $z$ of each
event in Eqs.~\ref{eq:lc}~and~\ref{eq:fz}. In order to avoid possible contributions
of the $\rho^0$ at high $z$ and large rescattering effects at low
$z$, only data with $0.3<z<0.7$ were used. The resultant values of
$\alpha$ and $\beta$ are shown in Fig.~\ref{fig:albe-lc}.
The behaviour of $\beta$ reflects the smaller attenuation at higher $L_c$
already visible in Fig.~\ref{fig:Rnuzlc}.
Consistent with what was found above, the value of
$\alpha$ decreases from about 0.6 at small $L_c$ to less than 0.5
at large $L_c$.

If, as was argued in the previous subsection, partonic effects
are most prominent at large values of $L_c$,
this would indicate that the $A$-dependence of that mechanism
has a value of $\alpha$ well below the value of 2/3 given in
Refs.~\cite{guo01,wang01} and \cite{arleo}.

The decrease of $\alpha$ with $L_c$ is not easily explained.
As demonstrated in Refs.~\cite{accardi,blo06} a pure absorption
mechanism would yield a value of $\alpha$ that increases with $L_c$.
Possibly such an increase is more than compensated for
by an increasing influence of the partonic mechanism.
However, then that mechamism should have an $A$-dependence with
a rather small value of $\alpha$.
But, when two (or possibly even more) mechanisms contribute,
the $A$-dependence probably becomes more complicated than can be
described by a single exponential as in Eq.~\ref{eq:adep}.
For that reason comparisons between data and model calculations
should be done on the level of the multiplcity ratio $R_A^h$.

\subsubsection{Other particles}

For the other particles the statistical precision is too low for a meaningful
two-dimensional binning and study of the $A$-dependence, so there the $A$-dependence
is presented for $\nu$-bins of $6.0-12.0-17.0-23.5$~GeV (integrated over
all $z$), and for $z$-bins of $0.2-0.4-0.7-1.2$ (integrated over all $\nu$).
For comparison the values for $\pi^+$ and $\pi^-$ in the same bins
are given as well.  The results are shown in Fig.~\ref{fig:albet_gar}.
The only significant feature is the different behaviour seen for
$K^+$ particles at high $\nu$. As mentioned before,
this may be due to large rescattering effects. The behaviour for
$K^-$ is different from the one for $K^+$, and more like that of
$\pi^-$, but with smaller values for $\alpha$.

\section{\label{sec:concl} Summary and conclusions}

Data for the multiplicity ratio $R_A^h$ of hadron production
in semi-inclusive deep-inelastic scattering of 27.6 GeV
electrons and positrons from
helium, neon, krypton, and xenon nuclear targets relative to
deuterium were obtained for identified
$\pi^+$, $\pi^-$, $\pi^0$, $K^+$,  $K^-$, $p$, and $\bar{p}$
particles as a function of the virtual-photon energy $\nu$,
the fraction $z$ of the energy transferred to the hadron,
the photon virtuality $Q^2$, and the hadron transverse
momentum squared $p_t^2$. For all particles the dependence
of $R_A^h$ on these variables is presented and discussed.

The most prominent features of the data are an increased
attenuation (decrease of $R_A^h$ below unity) with increasing value
of the mass number $A$ of the nucleus and the attenuation becoming
smaller (larger) with increasing values of $\nu$ ($z$),
$R_A^h$ dropping below 0.5 for xenon in some kinematic regions.
At low values of $z$, especially for heavier targets and for protons
and $K^+$ particles, a strong rise of $R_A^h$, even to above
unity, is observed. Presumably this is due to hadronic rescattering,
where a higher energy particle through nuclear reactions produces
one or more lower-energy particles.

The value of $R_A^h$ increases slightly with $Q^2$, at least for pions,
and is almost independent of $p_t^2$, except at large values
of $p_t^2$, where $R_A^h$ increases strongly.
The latter is thought to result from $p_t^2$ broadening due to partonic
rescattering (Cronin effect).
This effect was seen to disappear for $z \rightarrow 1$, in accordance
with the picture that in that limit no rescattering is possible,
since rescattering of the struck parton implies an energy loss.

By combining the data for $\pi^+$ and $\pi^-$, the dependence of
$R_A^{\pi}$ on two of the variables $\nu$, $z$, $Q^2$, and $p_t^2$
together was investigated.
The dependence on $Q^2$ depends weakly but noticeably on the value of $\nu$,
but practically not on that of $z$.
The dependence on $p_t^2$ hardly depends on $\nu$ and $z$,
except for the disapperance of the rise at large $p_t^2$ at
$z \rightarrow 1$ mentioned above.
However, the dependences on $\nu$ and $z$ are related.
It was found that most of the dependence on $\nu$ and $z$
can be incorporated in a dependence on the combination
$L_c=z^{0.35}(1-z)\nu/\kappa$,
where $\kappa$ is the string tension in string models,
which thus acts as a scaling variable.
Since this function is close to the one given in the Lund model for the
average formation length of a particle,
by inspecting the value of $R_A^\pi$~vs.~$L_c$ for the four nuclei,
regions can tentatively be identified, where hadronic (absorption)
plus partonic mechanisms are important, and a region at higher~$L_c$
where only or mainly partonic mechanisms play a role.

A fit of the $A$-dependence of the values of $R_A^\pi$ for pions 
measured for the various nuclei of the form
$R_{A}^\pi = \exp[-\beta (A/100)^\alpha]$
yields values of $\alpha$ from about 0.6 to 0.5, depending
on the value of $L_c$.
Similar values are found for the other particles.
These values of $\alpha$ are well below the values resulting
from models in which the attenuation depends on the square of
the  distance a parton travels through the nucleus.

In total a very extensive data set to guide modeling hadronization in
nuclear matter has been collected.
A full theoretical description of hadronization in nuclei in one
consistent framework, including partonic and hadronic
(absorption plus rescattering) mechanisms is badly needed.
Clearly it will be a challenge for any theoretical model that is
developed to describe these data for the various hadrons and
nuclei as a function of all kinematic variables, but
if successful, this combination of data and theoretical
interpretation will contribute essentially to the understanding
of non-perturbative QCD at normal, and thence higher densities.

We gratefully acknowledge the DESY management for its support and the staff
at DESY and the collaborating institutions for their significant effort.
This work was supported by the FWO-Flanders, Belgium;
the Natural Sciences and Engineering Research Council of Canada;
the National Natural Science Foundation of China;
the Alexander von Humboldt Stiftung;
the German Bundesministerium f\"ur Bildung und Forschung (BMBF); 
the Deutsche Forschungsgemeinschaft (DFG);
the Italian Istituto Nazionale di Fisica Nucleare (INFN);
the Monbusho International Scientific Research Program, JSPS,
and Toray Science Foundation of Japan;
the Dutch Foundation for Fundamenteel Onderzoek der Materie (FOM);
the U. K. Engineering and Physical Sciences Research Council, the
Particle Physics and Astronomy Research Council and the
Scottish Universities Physics Alliance;
the U. S. Department of Energy (DOE) and the National Science Foundation (NSF);
and the Ministry of Trade and Economical Development and the Ministry
of Education and Science of Armenia.

\bibliographystyle{./elsart-num}

\begin{appendix}
\section{Appendix}
\normalsize
In the next table the average values of the kinematic variables that
were integrated over when showing the dependences on $\nu$, $z$, $Q^2$
and $p_t^2$ in Figs.~\ref{fig:rplus},~\ref{fig:rmin},~\ref{fig:ptpn} are given for
the case of pion production on krypton. These values hardly depend on
the target used.

\begin{table}[ht]
\caption{\label{tab:avkin} 
Average values of $\nu$, $Q^2$, $z$, and $p_t^2$ for pions produced on
krypton.
}
\begin{center}
\renewcommand{\baselinestretch}{1.0}
\scriptsize
\begin{tabular}{ccccc}
$\nu$ range&$\langle \nu \rangle$ $(GeV)$&$\langle z
\rangle$&$\langle Q^2 \rangle$ $(GeV^2)$&$\langle 
p_t^2\rangle$ $(GeV^2)$\\
\hline
4.0 -  6.0& 5.269&0.583& 1.889& 0.245\\
6.0 -  8.0& 7.151&0.501& 2.029& 0.164\\
8.0 - 10.0& 9.092&0.448& 2.203& 0.141\\
10.0 - 12.0&11.030&0.421& 2.352& 0.147\\
12.0 - 14.0&13.009&0.406& 2.506& 0.171\\
14.0 - 16.0&14.991&0.392& 2.595& 0.199\\
16.0 - 18.0&16.980&0.370& 2.584& 0.234\\
18.0 - 20.0&18.964&0.349& 2.475& 0.275\\
20.0 - 23.5&21.540&0.326& 2.164& 0.334\\
$z$ range&$\langle z \rangle$&$\langle \nu \rangle$ 
$(GeV)$&$\langle
Q^2 \rangle$ $(GeV^2)$&$\langle 
p_t^2\rangle$ $(GeV^2)$\\
\hline
0.1 - 0.2& 0.155&18.358& 2.463& 0.104\\
0.2 - 0.3& 0.247&15.953& 2.623& 0.164\\
0.3 - 0.4& 0.345&14.796& 2.655& 0.219\\
0.4 - 0.5& 0.446&14.315& 2.635& 0.260\\
0.5 - 0.6& 0.546&13.963& 2.585& 0.284\\
0.6 - 0.7& 0.646&13.637& 2.500& 0.286\\
0.7 - 0.8& 0.746&12.824& 2.392& 0.246\\
0.8 - 0.9& 0.845&11.841& 2.252& 0.192\\
0.9 - 1.3& 0.952&10.742& 2.258& 0.156\\
$Q^2$ range&$\langle Q^2 \rangle$ $(GeV^2)$&$\langle \nu
\rangle$ $(GeV)$&$\langle z \rangle$&$\langle 
p_t^2\rangle$ $(GeV^2)$\\
\hline
1.0 - 1.5& 1.360&14.790& 0.399& 0.217\\
1.5 - 2.0& 1.911&14.646& 0.395& 0.212\\
2.0 - 3.0& 2.667&14.874& 0.389& 0.213\\
3.0 - 4.0& 3.746&15.062& 0.382& 0.213\\
4.0 - 5.0& 4.775&15.100& 0.380& 0.211\\
5.0 - 6.0& 5.798&15.065& 0.380& 0.208\\
6.0 - 8.0& 7.161&14.819& 0.380& 0.203\\
8.0 - 25.0& 9.735&14.583& 0.386& 0.192\\
$p_t^2$ range&$\langle p_t^2\rangle$ $(GeV^2)$&$\langle \nu
\rangle$ $(GeV)$&$\langle z \rangle$&$\langle Q^2 \rangle$ $(GeV^2)$\\
\hline
0.00 - 0.05& 0.023&13.059& 0.379& 2.550\\
0.05 - 0.10& 0.073&14.107& 0.390& 2.611\\
0.10 - 0.30& 0.182&15.276& 0.385& 2.599\\
0.30 - 0.50& 0.384&16.172& 0.392& 2.616\\
0.50 - 0.70& 0.585&16.669& 0.424& 2.635\\
0.70 - 0.90& 0.787&16.912& 0.456& 2.627\\
0.90 - 1.10& 0.987&17.211& 0.483& 2.620\\
1.10 - 1.40& 1.225&17.471& 0.505& 2.648\\
1.40 - 1.95& 1.606&17.788& 0.535& 2.615\\
1.95 - 5.00& 2.418&18.487& 0.571& 2.625\\
\end{tabular}
\end{center}
\end{table}
\end{appendix}
\end{document}